\documentclass[journal]{IEEEtran}

\ifCLASSINFOpdf
  \usepackage[pdftex]{graphicx}
  \DeclareGraphicsExtensions{.pdf,.jpeg,.png}
\else
  \usepackage[dvips]{graphicx}
\fi
\usepackage{epstopdf}
\usepackage[cmex10]{amsmath}
\usepackage{amssymb}
\usepackage{wasysym}
\usepackage{algorithm}
\usepackage{algorithmic}
\usepackage{array}
\usepackage{cite}
\usepackage{color}
\usepackage{url}

\usepackage{epsfig,latexsym}
\usepackage{flushend}
\usepackage{cite}
\usepackage{verbatim}
\usepackage{amsopn}
\usepackage{booktabs}

\usepackage{stfloats}

\begin{document}

\title{Beamspace Channel Estimation for Millimeter Wave Massive MIMO System with Hybrid Precoding and Combining}

\author{Wenyan~Ma,~\IEEEmembership{Student~Member,~IEEE}, Chenhao~Qi,~\IEEEmembership{Senior~Member,~IEEE}
\thanks{This work is supported in part by National Natural Science Foundation of China under Grant 61302097 and Natural Science Foundation of Jiangsu Province under Grant BK20161428. (\textit{Corresponding author: Chenhao~Qi})}
\thanks{Wenyan~Ma and Chenhao~Qi are with the School of Information Science and Engineering, Southeast University, Nanjing 210096, China (Email: qch@seu.edu.cn).}
}

\markboth{}
{Shell \MakeLowercase{\textit{et al.}}: Bare Demo of IEEEtran.cls for Journals}

\maketitle

\begin{abstract}
In this paper, a framework of beamspace channel estimation in millimeter wave (mmWave) massive MIMO system is proposed. The framework includes the design of hybrid precoding and combining matrix as well as the search method for the largest entry of over-sampled beamspace receiving matrix. Then based on the framework, three channel estimation schemes including identity matrix approximation (IA)-based scheme, scattered zero off-diagonal (SZO)-based scheme and concentrated zero off-diagonal (CZO)-based scheme are proposed. These schemes together with the existing channel estimation schemes are compared in terms of computational complexity, estimation error and total time slots for channel training. Simulation results show that the proposed schemes outperform the existing schemes and can approach the performance of the ideal case. In particular, total time slots for channel training can be substantially reduced.

\end{abstract}

\section{Introduction}
Millimeter wave (mmWave) communication is a promising technology for next generation wireless communication owing to its abundant frequency spectrum resource\cite{heath2016overview,li20155G}. However, realizing mmWave massive MIMO in practice is not a trivial task, which faces the problem of high propagation loss due to the high carrier frequency \cite{rappa2013work}. To compensate for the propagation loss, antenna arrays are usually used to form directional beamforming. Fortunately, thanks to the short wavelength of the mmWave frequency, large antenna arrays are possible to be packed into small form factors.

In order to exploit the spatial degree of freedom, the hybrid analog and digital precoding is usually employed \cite{alkhateeb2014mimo,ayach2014spatially,choi2017resolution}. A
small number of RF chains are tied to a large antenna array. This structure enables parallel transmission, and thus provides the potential to approach the capacity bound that can be achieved by digital precoding. On the other hand, the large antenna arrays challenge the low-complexity design of hybrid precoding and channel estimation \cite{yu2016alternating}. In particular, the hybrid precoding may require matrix operations with a scale of antenna size, which is generally large in mmWave communication \cite{alkhateeb2014mimo}. Moreover, the channel estimation is also rather time consuming due to the large number of antennas at both transmitting and receiving sides \cite{alkhateeb2014channel}.

To reduce the complexity of channel estimation in mmWave massive MIMO system, some advanced schemes based on the beamspace channel have been proposed very recently \cite{alkhateeb2014channel,xiao2016Hierarchical,chen2017compressive,li2015estimation,Kim2015virtual,yang2016efficient,gao2016estimation}. A hybrid and multiresolutional codebook (HMC)-based channel estimation method and a JOINT-based channel estimation method are proposed in \cite{alkhateeb2014channel} and \cite{xiao2016Hierarchical}, respectively. The common idea of \cite{alkhateeb2014channel} and \cite{xiao2016Hierarchical} is to use the hierarchical codebook, where the precision of channel estimation relies on the number of the layers in the hierarchical codebook. The key ideas of \cite{li2015estimation,Kim2015virtual,yang2016efficient,gao2016estimation} are to efficiently explore the sparsity of beamspace channel by sparse signal processing techniques. An enhanced compressive sensing (ECS)-based channel estimation scheme is proposed in \cite{li2015estimation}, which explores the channel sparsity in beamspace. With higher resolution of phase shifters, improved channel estimation accuracy can be achieved since higher freedom is available for the design of measurement matrix for the sparse recovery\cite{alkhateeb2014mimo,ayach2014spatially}.

However, considering the limited beamspace resolution, the sparsity of beamspace channel may be impaired by power leakage \cite{dai2016estimation}, indicating that the beamspace channel is not ideally sparse and there are many small nonzero entries \cite{dai2017tras}. Therefore, it brings extra challenge for the sparse recovery \cite{alkhateeb2015compressed,tropp2007omp}. To solve this problem, an adaptive support detection (ASD)-based channel estimation scheme is proposed to iteratively detect and adjust the channel support to find one with the largest channel power \cite{gao2016estimation}. Note that the beamspace resolution is proportional to the reciprocal of the antenna number.  Therefore, ASD-based scheme cannot obtain high precision channel estimation regarding the channel power leakage \cite{alkhateeb2014channel}. A discrete compressive sensing (DCS)-based channel estimation scheme is proposed in \cite{dai2016estimation}, where the beamspace resolution can be set arbitrarily and can be higher than the reciprocal of the antenna number, leading to better channel estimation performance than ASD-based scheme. An over-sampling compressive sensing (OCS)-based channel estimation scheme is proposed in \cite{alkhateeb2015compressed}, where the measurement matrix is consisted of a large number of over-sampling steering vectors and is capable of estimating the angle of arrival (AoA) and angle of departure (AoD) more accurate than conventional measurement matrix. However, both DCS-based and OCS-based schemes use random measurement matrices, which can not always achieve the optimal performance. In particular, the estimated AoA and AoD can not be always within the scale of quantization error even without any noise. Therefore, it is better to use deterministic measurement matrix.

In this paper, we first propose a framework of beamspace channel estimation, which is divided into two subproblems, including the design of the hybrid precoding and combining matrix, and the search method for the largest entry of over-sampled beamspace receiving matrix. Note that the design of the hybrid precoding and combining matrix as well as the search method for the largest entry is original. Then based on the framework, we propose three channel estimation schemes.

1) We propose an identity matrix approximation (IA)-based channel estimation scheme. We formulate the design of hybrid combining and hybrid precoding as two optimization problems with the constraint of total power and constraint of the constant envelope required by phase shifters. Due to the non-convexity of the problems, we decouple the design of analog combining and digital combining, and then obtain closed-form solutions. We propose an algorithm for the design of hybrid combining matrix. The algorithm repeatedly fixes the analog combining matrix to obtain digital combining matrix, and then fixes the digital combining matrix to obtain the analog combining matrix in turn, until the stop condition is satisfied. Detailed steps summarized in an algorithm table are provided. Since the design of the hybrid precoding matrix is similar, we briefly describe the design of the hybrid precoding matrix. Note that the hybrid combining matrix and the hybrid precoding matrix can be designed off-line before the channel training. After that, we propose an algorithm to search the largest entry of the over-sampled beamspace receiving matrix. The algorithm is based on trichotomy search and includes two stages, where we find the main lobe in the first stage, and we find the largest entry corresponding to the peak within the main lobe in the second stage.

2) We design hybrid precoding matrix and hybrid combining matrix so that the coordinates of the largest entry of over-sampled beamspace receiving matrix are the AoA and AoD within the quantization error. Then we convert the discrete problem into a continuous problem to obtain the derivative. By setting the derivative zero, we get two solutions where the first solution is found to be meaningless. Based on the second solution, we propose two zero off-diagonal (ZO) beamspace channel estimation schemes, namely, scattered zero off-diagonal (SZO)-based scheme and concentrated zero off-diagonal (CZO)-based scheme.
\begin{itemize}
    \item In the SZO-based channel estimation scheme, the nonzero diagonal entries are uniformly distributed with the same interval. Since the main lobe and the side lobes have the same envelope, we first make beam training based on codebook to find the main lobe and then use complementary channel estimation to further estimate the channel AoA and AoD within the main lobe. An integration-based codebook design method which results in closed-form expression of codewords is proposed and compared with the existing sparse-based codebook design method. Note that the proposed integration-based codebook design method is original.
    \item In the CZO-based channel estimation scheme, the nonzero diagonal entries are concentrated on the upper left corner of the matrix. Since the envelope of the main lobe and the side lobe is different, we can directly employ the algorithm proposed in the IA-based scheme to search the largest entry of over-sampled beamspace receiving matrix corresponding to the channel AoA and AoD within the main lobe.

    \end{itemize}

Additionally, we also compare the above three schemes together with the existing channel estimation schemes in terms of computational complexity, estimation error and total time slots for channel training.

The rest of the paper is organized as follows. In Section II, the system model and problem formulation of beamspace channel estimation with hybrid precoding and combining are provided. In Section~III, we propose three beamspace channel estimation schemes. The simulation results are provided in Section IV. Finally, Section V concludes the paper.

The notations are defined as follows. Symbols for matrices (upper case) and vectors (lower case) are in boldface. $(\cdot)^T $, $(\cdot)^H $, $(\cdot)^{-1} $, $\boldsymbol{I}_{L}$, $\textbf{1}_{L}$, $\mathbb{C}^{M\times{N}}$, $\otimes$, $\textrm{vec}(\cdot)$, $\operatorname{E}\{\cdot\}$, $\mathcal{O}(\cdot)$, $\boldsymbol{0}^M$, $\boldsymbol{0}_{M\times N}$, $\|\cdot\|_0$, $\|\cdot\|_2$, $\|\cdot\|_F$, $\boldsymbol{A}[p,q]$, $\langle\cdot\rangle$, $\lfloor\cdot\rfloor$, $\textrm{Tr}(\cdot)$, $\mathbb{Z}$, $\mathbb{Re}\{\cdot\}$ and $\mathcal{CN}$, denote the transpose, conjugate transpose (Hermitian), inverse, identity matrix of size $L$, vector of size $L$ with all entries being 1, the set of $M\times{N}$ complex-valued matrices, kronecker product, vectorization, expectation, order of complexity, zero vector of size $M$, $M\times{N}$ zero matrix, $l_0$-norm, $l_2$-norm, Frobenius norm, entry of $\boldsymbol{A}$ at the $p$th row and $q$th column, round function, floor function, trace, set of integer, real part and complex Gaussian distribution, respectively.

\section{System Model and Problem Formulation}
\subsection{System Model}
We consider a time division duplexing (TDD) multi-user mmWave massive MIMO system comprising a base station (BS) and $U$ users. We focus on the uplink transmission. Both the BS and users are equipped with an uniform linear array (ULA)~\cite{amadori2015low}. Let $N_A$, $M_A$, $N_R$ and $M_R$ denote the number of antennas at the BS, number of antennas at each user, number of RF chains at the BS and number of RF chains at each user. In practical mmWave massive MIMO with hybrid precoding and combining, the number of RF chains is much smaller than that of antennas, i.e., $N_R \ll N_A$ and $M_R \ll M_A$. This is because we use large antenna array to form
directional beamforming, which can compensate for the propagation loss caused by the high carrier frequency \cite{gao2016near}.

For uplink transmission, each user performs analog precoding in RF and digital precoding in the baseband, while the BS performs analog combining in RF and digital combining in the baseband. The received signal vector at the BS can be represented as
\begin{equation}\label{UpSig}
\boldsymbol{y} = \boldsymbol{W}_B \boldsymbol{W}_R  \sum_{u=1}^{U} \boldsymbol{H}_u \boldsymbol{F}_{R,u} \boldsymbol{F}_{B,u} \boldsymbol{s}_u + \boldsymbol{W}_B \boldsymbol{W}_R \boldsymbol{n}
\end{equation}
where $\boldsymbol{F}_{B,u}\in{\mathbb{C}^{M_R\times{M_R}}}$, $\boldsymbol{F}_{R,u}\in{\mathbb{C}^{M_A\times{M_R}}}$, $\boldsymbol{W}_B\in{\mathbb{C}^{N_R\times{N_R}}}$, and $\boldsymbol{W}_R\in{\mathbb{C}^{N_R\times{N_A}}}$ are the digital precoding matrix, analog precoding matrix, digital combining matrix, and analog combining matrix for the $u(u=1,2,\ldots,U)$th user, respectively. $\boldsymbol{s}_u\in{\mathbb{C}^{M_R}}$ denotes the signal vector satisfying $\operatorname{E}\{\boldsymbol{s}_u\boldsymbol{s}_u^{H}\}=\boldsymbol{I}_{M_R}$. $\boldsymbol{n} \in{\mathbb{C}^{N_A}} $ denotes additive white Gaussian noise (AWGN) vector satisfying $\boldsymbol{n} \sim \mathcal{CN}(0,\sigma^{2}\boldsymbol{I}_{N_A})$. $\boldsymbol{H}_u\in{\mathbb{C}^{N_A\times{M_A}}}$ denotes the channel matrix between the BS and the $u$th user and can be expressed according to the widely used Saleh-Valenzuela channel
model \cite{heath2016overview} as
\begin{equation}\label{ULAchannelmodel}
\boldsymbol{H}_{u}=\sqrt{\frac{N_AM_A}{L_u}}\sum_{i=1}^{L_u} g_{u,i} \boldsymbol{\alpha} (N_A,\theta_{u,i}) \boldsymbol{\alpha}^{H} (M_A,\varphi_{u,i})
\end{equation}
where $L_{u}$ and $g_{u,i}$ denote the total number of resolvable paths and the channel fading coefficient of the $i$th path for the $u$th user, respectively. Usually, there is a strong line-of-sight (LOS) path ($i=1$) and $L_u-1$ much weaker non-line-of-sight (NLOS) paths ($2\leq i \leq L_u$). The mmWave transmission essentially relies on the strong LOS path. The steering vector $\boldsymbol{\alpha}(N,\theta)$ is defined as
\begin{equation}
\boldsymbol{\alpha}(N,\theta)=\frac{1}{\sqrt{N}}\left[1,e^{-j\pi\theta},...,e^{-j\pi\theta(N-1)}\right]^{T}.
\end{equation}
Define the AoA and AoD of the $i$th path of the $u$th user as $\Theta_{u,i}$ and $\Phi_{u,i}$, respectively. Further define $\theta_{u,i} \triangleq \frac{2d_{BS}}{\lambda}\sin{\Theta_{u,i}}$ and $\varphi_{u,i} \triangleq \frac{2d_{UE}}{\lambda}\sin{\Phi_{u,i}}$, where $d_{BS}$ and $d_{UE}$ denote the antenna interval of the BS and users, respectively. We usually set $d_{BS}=d_{UE}=\lambda/2$, where $\lambda$ is the wavelength of mmWave signal. In practice, both $\Theta_{u,i}$ and $\Phi_{u,i}$ obey the uniform distribution $[-\pi,\pi]$ \cite{alkhateeb2014channel,rappaport2015wideband}.

\subsection{Problem Formulation}
Note that $\boldsymbol{y}$ in (\ref{UpSig}) is a combination of signal from different users. We use $T_1$ different digital precoding matrices and analog precoding matrices, denoted as $\boldsymbol{F}_{B,u}^{t_1} \in{\mathbb{C}^{M_R\times{M_R}}} $ and $\boldsymbol{F}_{R,u}^{t_1}\in{\mathbb{C}^{M_A\times{M_R}}}$, respectively, $t_1=1,2,\ldots,T_1$, at the $u(u=1,2,\ldots,U)$th user. We use $T_2$ different digital combining matrices and analog combining matrices, denoted as $\boldsymbol{W}_{B}^{t_2}\in{\mathbb{C}^{N_R\times{N_R}}}$ and $\boldsymbol{W}_{R}^{t_2}\in{\mathbb{C}^{N_R\times{N_A}}}$, respectively, $t_2=1,2,\ldots,T_2$, at the BS. To distinguish different user signal at the BS, each user repeatedly transmits an orthogonal pilot sequence $\boldsymbol{p}_u \in{\mathbb{C}^U} $ for $T_1 T_2$ times. For simplicity, we suppose each user transmit the same pilot sequence for all $M_R$ RF chains, where the pilot matrix for the $u$th user can be defined as $\boldsymbol{P}_u \triangleq [\boldsymbol{p}_u,\boldsymbol{p}_u,\ldots,\boldsymbol{p}_u]^H = \textbf{1}_{M_R}\boldsymbol{p}_u^H \in{\mathbb{C}^{M_R\times{U}}}$. The channel keeps constant during $T \triangleq T_1 T_2 U$ time slots \cite{dai2017tras}. During the $T_1$ repetitive transmission of pilot sequence from the $((t_2-1)T_1+1)$th transmission to $(t_2 T_1)$th transmission, we use $T_1$ different $\boldsymbol{F}_{B,u}^{t_1}$ and $\boldsymbol{F}_{R,u}^{t_1}$ for hybrid precoding while using the same $\boldsymbol{W}_{B}^{t_2}$ and $\boldsymbol{W}_{R}^{t_2}$ for hybrid combining, where the received pilot matrix $\boldsymbol{Y}^{t_1,t_2}\in{\mathbb{C}^{N_R\times{U}}}$ can be denoted as
\begin{equation}
\boldsymbol{Y}^{t_1,t_2} = \boldsymbol{W}_B^{t_2} \boldsymbol{W}_R^{t_2}  \sum_{u=1}^{U} \boldsymbol{H}_u \boldsymbol{F}_{R,u}^{t_1} \boldsymbol{F}_{B,u}^{t_1} \boldsymbol{P}_u  +  \boldsymbol{W}_B^{t_2} \boldsymbol{W}_R^{t_2} \boldsymbol{N}^{t_1,t_2}
\end{equation}
with $\boldsymbol{N}^{t_1,t_2}\in{\mathbb{C}^{N_A\times{U}}}$ representing the AWGN matrix. Each entry of $\boldsymbol{N}^{t_1,t_2}$ independently obeys complex Gaussian distribution with zero mean and variance of $\sigma^2$. To ease the notation, we define $\widetilde{\boldsymbol{N}}^{t_1,t_2} \triangleq \boldsymbol{W}_B^{t_1} \boldsymbol{W}_R^{t_1} \boldsymbol{N}^{t_1,t_2}$. Due to the orthogonality of $\boldsymbol{p}_u$, i.e., $\boldsymbol{p}_u^H$$\boldsymbol{p}_u$=1 and $\boldsymbol{p}_u^H$$\boldsymbol{p}_i=0$, $\forall u,i\in \{1,2,\ldots,U\}$, $i\neq{u}$ \cite{gao2016estimation}, we can obtain the measurement vector $\boldsymbol{r}_u^{t_1,t_2}\in{\mathbb{C}^{N_R}}$ for the $u$th user by multiplying $\boldsymbol{Y}^{t_1,t_2}$ with $\boldsymbol{p}_u$ as
\begin{equation}
\boldsymbol{r}_u^{t_1,t_2} = \boldsymbol{Y}^{t_1,t_2} \boldsymbol{p}_u = \boldsymbol{W}^{t_2} \boldsymbol{H}_u \boldsymbol{f}^{t_1}_u + \widetilde{\boldsymbol{n}}^{t_1,t_2}
\end{equation}
where
\begin{equation}\label{Wt2}
\begin{split}
\boldsymbol{W}^{t_2}  \triangleq \boldsymbol{W}_B^{t_2}\boldsymbol{W}_R^{t_2},  &  ~~\boldsymbol{f}^{t_1}_u  \triangleq{\boldsymbol{F}_{R,u}^{t_1} \boldsymbol{F}_{B,u}^{t_1} \textbf{1}_{M_R}},\\
\widetilde{\boldsymbol{n}}^{t_1,t_2} &\triangleq \widetilde{\boldsymbol{N}}^{t_1,t_2} \boldsymbol{p}_u.
\end{split}
\end{equation}
Note that we can distinguish different user signal utilizing the orthogonality of pilot sequences. In this way, the spatial distribution of users has no impact on the performance of our proposed schemes. Define $T_3 \triangleq T_2N_R$. We stack the $T_2$ received pilot sequences together and have
\begin{equation}
\boldsymbol{r}_u^{t_1} = \boldsymbol{W} \boldsymbol{H}_u \boldsymbol{f}^{t_1}_u + \widetilde{\boldsymbol{n}}^{t_1}
\end{equation}
where
\begin{equation}\label{DefinitionW}
\begin{split}
 \boldsymbol{r}_u^{t_1} &\triangleq [(\boldsymbol{r}_u^{t_1,1})^T , (\boldsymbol{r}_u^{t_1,2})^T , \ldots , (\boldsymbol{r}_u^{t_1,T_2})^T]^T \in{\mathbb{C}^{T_3}}, \\
 \boldsymbol{W} &\triangleq [(\boldsymbol{W}^{1})^T , (\boldsymbol{W}^{2})^T , \ldots , (\boldsymbol{W}^{T_2})^T]^T \in{\mathbb{C}^{T_3 \times{N_A}}}, \\
 \widetilde{\boldsymbol{n}}^{t_1} &\triangleq [(\widetilde{\boldsymbol{n}}^{t_1,1})^T , (\widetilde{\boldsymbol{n}}^{t_1,2})^T , \ldots , (\widetilde{\boldsymbol{n}}^{t_1,T_2})^T]^T \in{\mathbb{C}^{T_3}}.\\
\end{split}
\end{equation}
Further define
\begin{equation}\label{DefinitionF}
\begin{split}
 \boldsymbol{R}_u &\triangleq [\boldsymbol{r}_u^{1} , \boldsymbol{r}_u^{2} , \ldots , \boldsymbol{r}_u^{T_1}] \in{\mathbb{C}^{T_3 \times T_1}}, \\
 \boldsymbol{F}_u &\triangleq [\boldsymbol{f}^{1}_u , \boldsymbol{f}^{2}_u , \ldots , \boldsymbol{f}^{T_1}_u] \in{\mathbb{C}^{M_A \times T_1}}, \\
 \widetilde{\boldsymbol{n}} &\triangleq [\widetilde{\boldsymbol{n}}^{1} , \widetilde{\boldsymbol{n}}^{2} , \ldots, \widetilde{\boldsymbol{n}}^{T_1}] \in{\mathbb{C}^{T_3 \times T_1}}.\\
\end{split}
\end{equation}
We have
\begin{equation}\label{Ru}
\boldsymbol{R}_u = \boldsymbol{W} \boldsymbol{H}_u \boldsymbol{F}_u + \widetilde{\boldsymbol{n}}.
\end{equation}
Considering the LOS channel path concentrates most channel power in mmWave massive MIMO systems, we usually use LOS channel path to transmit data for the $u$th user \cite{dai2016estimation}. Therefore, it is important to design $\boldsymbol{W}$ and $\boldsymbol{F}_u$ to estimate the AoA and the AoD of the LOS path of $\boldsymbol{H}_u$ in (\ref{Ru}) for the $u$th user, which will be discussed in the following sections.

\section{Beamspace Channel Estimation}
In this section, we first propose a framework of beamspace channel estimation. Then based on this framework, three channel estimation schemes are proposed. Finally, the comparisons of these three schemes together with the existing HMC-based \cite{alkhateeb2014channel}, JOINT-based \cite{xiao2016Hierarchical}, ECS-based \cite{li2015estimation}, DCS-based \cite{dai2016estimation} and OCS-based \cite{alkhateeb2015compressed} channel estimation schemes are also presented.

\begin{figure*}[ht]
\begin{equation} \label{ObjSimplify}
  \begin{split}
  &\| \boldsymbol{D}(N_A,K)^H \boldsymbol{W}^H \boldsymbol{W} - \gamma_N \boldsymbol{D}(N_A,K)^H \| _F^2 = \| \boldsymbol{D}(N_A,K)^H ( \boldsymbol{W}^H \boldsymbol{W} - \gamma_N \boldsymbol{I}_{N_A}) \| _F^2   \\
  = & \textrm{Tr} \big( (   \boldsymbol{W}^H \boldsymbol{W} - \gamma_N \boldsymbol{I}_{N_A})^H \boldsymbol{D}(N_A,K) \boldsymbol{D}(N_A,K)^H  (\boldsymbol{W}^H\boldsymbol{W} - \gamma_N \boldsymbol{I}_{N_A})  \big)  \\
  = & K \textrm{Tr} \big( (   \boldsymbol{W}^H \boldsymbol{W} - \gamma_N \boldsymbol{I}_{N_A})^H   (\boldsymbol{W}^H\boldsymbol{W} - \gamma_N \boldsymbol{I}_{N_A})  \big) / N_A = K \| \boldsymbol{W}^H \boldsymbol{W} - \gamma_N \boldsymbol{I}_{N_A} \|_F^2 / N_A
\end{split}
\end{equation}
\hrulefill
\end{figure*}

\subsection{Framework of Beamspace Channel Estimation}
The beamspace channel matrix for the $u$th user $\bar{\boldsymbol{H}}_u^v \in{\mathbb{C}^{N_A \times{M_A}}}$ can be represented as \cite{gao2016estimation}
\begin{equation}\label{bar_Huv}
\bar{\boldsymbol{H}}_u^v = \boldsymbol{D}(N_A,N_A)^H \boldsymbol{H}_u \boldsymbol{D}(M_A,M_A)
\end{equation}
where $\boldsymbol{D}(N,K) \in{\mathbb{C}^{N \times{K}}}$ is the sampling matrix, which is defined as
\begin{align}\label{OversmaplingDmatrix}
  \boldsymbol{D}(& N, K) \triangleq  [\boldsymbol{\alpha}(N,-1+0/K) , \boldsymbol{\alpha}(N,-1+2/K), \notag \\ & \boldsymbol{\alpha}(N,-1+4/K) , \ldots , \boldsymbol{\alpha}(N,-1+2(K-1)/K)].
\end{align}
In fact, $\boldsymbol{D}(N,K)$ samples the beamspace [-1,1] in an interval of $2/K$ by $K$ steering vectors. For $\bar{\boldsymbol{H}}_u^v$, the AoA and AoD is sampled in an interval of $2/N_A$ and $2/M_A$, respectively. Therefore the quantization error for the estimated AoA and AoD is $2/N_A$ and $2/M_A$, respectively. In order to decrease the quantization error, we introduce the over-sampled beamspace channel matrix for the $u$th user $\boldsymbol{H}_u^v \in{\mathbb{C}^{K \times{K}}}$ as
\begin{equation}\label{Huv}
\boldsymbol{H}_u^v = \boldsymbol{D}(N_A,K)^H \boldsymbol{H}_u \boldsymbol{D}(M_A,K)
\end{equation}
where $K$ is the number of steering vectors with $K>N_A$ and $K>M_A$. Then the coordinates of the largest entry of $\boldsymbol{H}_u^v$ are the AoA and AoD of the LOS path with the quantization error of $2/K$. To reduce the quantization error and improve channel estimation,  we can use a large $K$ by finding the largest entry of $\boldsymbol{H}_u^v$.

However, we cannot directly obtain $\boldsymbol{H}_u^v$ based on $\boldsymbol{R}_u$ in (\ref{Ru}), due to the hybrid precoding and combining operations of $\boldsymbol{F}_u$ and $\boldsymbol{W}$, respectively. Note that the dimension of $\boldsymbol{H}_u$ is $N_A \times M_A$, while the dimension of $\boldsymbol{W} \boldsymbol{H}_u \boldsymbol{F}_u$ is $T_3 \times T_1$. In order to obtain the over-sampled beamspace channel matrix as described in (\ref{Huv}), we multiply $\boldsymbol{R}_u$ with $\boldsymbol{W}^H$ on the left and $\boldsymbol{F}_u^H$ on the right, which can make  the dimension of $\boldsymbol{W}^H \boldsymbol{W} \boldsymbol{H}_u \boldsymbol{F}_u \boldsymbol{F}_u^H$ the same as that of $\boldsymbol{H}_u$. Now we can obtain an over-sampled beamspace receiving matrix $\boldsymbol{R}_u^v \in{\mathbb{C}^{K \times{K}}}$ as
\begin{equation}\label{Ruv}
\boldsymbol{R}_u^v = \boldsymbol{D}(N_A,K)^H \boldsymbol{W}^H \boldsymbol{W} \boldsymbol{H}_u \boldsymbol{F}_u \boldsymbol{F}_u^H \boldsymbol{D}(M_A,K) + \widetilde{\boldsymbol{n}}^v
\end{equation}
where $\widetilde{\boldsymbol{n}}^v \triangleq \boldsymbol{D}(N_A,K)^H \boldsymbol{W}^H \widetilde{\boldsymbol{n}} \boldsymbol{F}_u^H \boldsymbol{D}(M_A,K)$. It is expected that
\begin{equation}\label{IdealWF}
\begin{split}
\boldsymbol{D}(N_A,K)^H \boldsymbol{W}^H \boldsymbol{W}&=\gamma_N \boldsymbol{D}(N_A,K)^H, \\
\boldsymbol{F}_u \boldsymbol{F}_u^H \boldsymbol{D}(M_A,K) &= \gamma_M \boldsymbol{D}(M_A,K).
\end{split}
\end{equation}
where $\gamma_N \triangleq \|\boldsymbol{W}^H\boldsymbol{W}\|_F / \sqrt{N_A}$ and $\gamma_M \triangleq \|\boldsymbol{F}_u\boldsymbol{F}_u^H\|_F / \sqrt{M_A}$. However, in this case, it requires that $T_3\geq N_A$ and $T_1 \geq M_A$, leading to huge pilot overhead. For example, in an mmWave massive MIMO system with $N_A=64$, $M_A=16$, $N_R=4$ and $M_R=1$, the pilot sequence should be repetitively transmitted for $T_1 T_2=256$ times. Therefore, it is required to reduce the pilot overhead in practice, where none of $\boldsymbol{D}(N_A,K)^H \boldsymbol{W}^H \boldsymbol{W}=\gamma_N \boldsymbol{D}(N_A,K)^H$ and $\boldsymbol{F}_u \boldsymbol{F}_u^H \boldsymbol{D}(M_A,K) = \gamma_M \boldsymbol{D}(M_A,K)$ can be satisfied. Now we have the following two subproblems.

\emph{Subproblem 1 (Hybrid Precoding and Combining Matrix Design):} $\boldsymbol{W}$ and $\boldsymbol{F}_u$ should be well designed so that the coordinates of the largest entry of $\boldsymbol{R}_u^v$ are the AoA and AoD of the LOS path with the quantization error of $2/K$. Therefore, the AoA and AoD of the LOS path can be estimated by finding the largest entry of $\boldsymbol{R}_u^v$.

\emph{Subproblem 2 (Search the Largest Entry):} Due to the large dimension of $\boldsymbol{R}_u^v$, finding the largest entry of $\boldsymbol{R}_u^v$ is computationally expensive. Therefore, it is better to design a low-complexity search algorithm fully regarding the structure of $\boldsymbol{R}_u^v$.

\subsection{IA-based Beamspace Channel Estimation Scheme}
\subsubsection{Hybrid Precoding and Combining Matrix Design}
It is observed that $\boldsymbol{R}_u^v$ in (\ref{Ruv}) is the over-sampled beamspace channel matrix with noise if (\ref{IdealWF}) is satisfied. However, due to the fact that $T_3<N_A$, $T_1<M_A$, the rank of $\boldsymbol{W}^H \boldsymbol{W}$ and $\boldsymbol{F}_u \boldsymbol{F}_u^H$ is less than $N_A$ and $M_A$, respectively. So we cannot find a proper $\boldsymbol{W}$ and $\boldsymbol{F}_u$ satisfying (\ref{IdealWF}).

The optimization problem for \textbf{hybrid combining matrix design} is
\begin{equation}\label{minW}
\begin{split}
\underset{ \boldsymbol{W}} {\bf min}  \quad & \| \boldsymbol{D}(N_A,K)^H \boldsymbol{W}^H \boldsymbol{W} - \gamma_N \boldsymbol{D}(N_A,K)^H \| _F \\
{\bf s.t.} \quad & \boldsymbol{W}_R^{t_2} \in \mathcal{W_R},~t_2 = 1,2,\ldots,T_2, \\
                 & \| \boldsymbol{W}_B^{t_2} \boldsymbol{W}_R^{t_2} \| _F^2 = P_W, ~t_2 = 1,2,\ldots,T_2, \\
\end{split}
\end{equation}
where $\mathcal{W_R}$ is the set of all feasible analog combining matrix and $P_W=1$ to normalize the hybrid combining matrix. Note that the design of hybrid precoder and hybrid combiner is independent. We can well design the hybrid precoding and combining matrices before the transmission of pilot sequences. Note that $\boldsymbol{D}(N_A,K) \boldsymbol{D}(N_A,K)^H=K\boldsymbol{I}_{N_A}/N_A$. We have~(\ref{ObjSimplify}).
Then (\ref{minW}) can be further rewritten as
\begin{equation}\label{minW2}
\begin{split}
\underset{ \boldsymbol{W}} {\bf min}  \quad & \| \boldsymbol{W}^H \boldsymbol{W} - \gamma_N \boldsymbol{I}_{N_A} \| _F \\
{\bf s.t.} \quad & \boldsymbol{W}_R^{t_2} \in \mathcal{W_R},~t_2 = 1,2,\ldots,T_2, \\
                 & \| \boldsymbol{W}_B^{t_2} \boldsymbol{W}_R^{t_2} \| _F^2 = P_W, ~t_2 = 1,2,\ldots,T_2. \\
\end{split}
\end{equation}
It is seen that $\boldsymbol{W}$ defined in (\ref{DefinitionW}) is a flat matrix where the columns are more than the rows. Therefore, it is infeasible that $\boldsymbol{W}^H \boldsymbol{W}$ equals $\gamma_N \boldsymbol{I}_{N_A}$. Instead, it is important to design $\boldsymbol{W}$ so that $\boldsymbol{W}^H \boldsymbol{W} / \gamma_N$ approximates the identity matrix, a.k.a, identity matrix approximation (IA). To minimize $ \| \boldsymbol{W}^H \boldsymbol{W} - \gamma_N \boldsymbol{I}_{N_A} \| _F$, $\boldsymbol{W}$ can be a submatrix of $\sqrt{\gamma_N}\boldsymbol{U}$ by selecting the first $T_3$ rows of $\sqrt{\gamma_N}\boldsymbol{U}$, where $\boldsymbol{U}$ is any $N_A \times N_A$ unitary matrix \cite{jennings1992matrix}. For example, we obtain $\boldsymbol{U}$ by singular value decomposition (SVD) of a $N_A \times N_A$ random matrix $\boldsymbol{A}$, i.e., $\boldsymbol{A}=\boldsymbol{U\Sigma V}^H$, where each entry of $\boldsymbol{A}$ obeys the uniform distribution [0,1]. In this way, we obtain $\widetilde{\boldsymbol{W}}$. According to ($\ref{DefinitionW}$), we can obtain $\widetilde{\boldsymbol{W}}^{t_2},~t_2=1,2,\ldots,T_2$, which is essentially dividing $\widetilde{\boldsymbol{W}}$ into $T_2$ submatrices. Then (\ref{minW2}) is converted into $T_2$ subproblems, where each subproblem can be expressed as
\begin{equation}\label{minW3}
\begin{split}
\underset{ \boldsymbol{W}_B^{t_2},\boldsymbol{W}_R^{t_2}}{\bf min}  \quad & \| \boldsymbol{W}_B^{t_2} \boldsymbol{W}_R^{t_2} - \widetilde{\boldsymbol{W}}^{t_2} \| _F \\
{\bf s.t.} \quad & \boldsymbol{W}_R^{t_2} \in \mathcal{W_R},~\| \boldsymbol{W}_B^{t_2} \boldsymbol{W}_R^{t_2} \| _F^2 = P_W. \\
\end{split}
\end{equation}
We cannot directly obtain solutions for (\ref{minW3}) due to the non-convexity of the constraints. Note that \cite{yu2015hybrid} has proved that temporarily neglecting the second power constraint of (\ref{minW3}) during the optimization of $\boldsymbol{W}_B^{t_2}$ and $\boldsymbol{W}_R^{t_2}$  will have little impact on the optimality of the hybrid precoding problem. After $\boldsymbol{W}_B^{t_2}$ and $\boldsymbol{W}_R^{t_2}$ are obtained, we may set $\boldsymbol{W}_B^{t_2}$ as
\begin{equation}\label{Alg1SecondConstraintAdjustment}
{\boldsymbol{W}}_B^{t_2} \leftarrow \sqrt{P_W} \frac{{\boldsymbol{W}}_B^{t_2}}{ \| {\boldsymbol{W}}_B^{t_2} {\boldsymbol{W}}_R^{t_2} \| _F},~t_2= 1,2,\ldots,T_2
\end{equation}
to satisfy the second constraint of (\ref{minW3}).

To mitigate the
interference among different data streams, we impose a common constraint that the columns of the digital combining matrix are mutually orthogonal, i.e., ${\boldsymbol{W}_B^{t_2}}^H \boldsymbol{W}_B^{t_2} = \beta \boldsymbol{I}_{N_R}$, where $\beta \triangleq P_W/(N_AN_R)$. Define $\boldsymbol{W}_D^{t_2} \triangleq \beta^{-1}\boldsymbol{W}_B^{t_2}$. Then we have
\begin{align}
   & \| \boldsymbol{W}_B^{t_2} \boldsymbol{W}_R^{t_2} - \widetilde{\boldsymbol{W}}^{t_2} \| _F^2 \notag \\
   &= \| \boldsymbol{W}_B^{t_2} (\boldsymbol{W}_R^{t_2} - \beta^{-1}{\boldsymbol{W}_B^{t_2}}^H \widetilde{\boldsymbol{W}}^{t_2}) \| _F^2  \notag  \\
   &= \textrm{Tr} \big( (\boldsymbol{W}_R^{t_2} - {\boldsymbol{W}_D^{t_2}}^H \widetilde{\boldsymbol{W}}^{t_2})^H {\boldsymbol{W}_B^{t_2}}^H \boldsymbol{W}_B^{t_2} (\boldsymbol{W}_R^{t_2} - {\boldsymbol{W}_D^{t_2}}^H \widetilde{\boldsymbol{W}}^{t_2}) \big) \notag  \\
   &= \textrm{Tr} \big( (\boldsymbol{W}_R^{t_2} - {\boldsymbol{W}_D^{t_2}}^H \widetilde{\boldsymbol{W}}^{t_2})^H \beta (\boldsymbol{W}_R^{t_2} - {\boldsymbol{W}_D^{t_2}}^H \widetilde{\boldsymbol{W}}^{t_2}) \big) \notag  \\
   &= \beta \textrm{Tr} \big( (\boldsymbol{W}_R^{t_2} - {\boldsymbol{W}_D^{t_2}}^H \widetilde{\boldsymbol{W}}^{t_2})^H (\boldsymbol{W}_R^{t_2} - {\boldsymbol{W}_D^{t_2}}^H \widetilde{\boldsymbol{W}}^{t_2}) \big)  \notag  \\
   &= \beta \| \boldsymbol{W}_R^{t_2} - {\boldsymbol{W}_D^{t_2}}^H \widetilde{\boldsymbol{W}}^{t_2} \| _F^2.
\end{align}
Therefore (\ref{minW3}) can be expressed as
\begin{equation}\label{minW32}
\begin{split}
\underset{ \boldsymbol{W}_B^{t_2},\boldsymbol{W}_R^{t_2}} {\bf min}  \quad & \| \boldsymbol{W}_R^{t_2} - {\boldsymbol{W}_D^{t_2}}^H \widetilde{\boldsymbol{W}}^{t_2} \| _F^2 \\
{\bf s.t.} \quad & \boldsymbol{W}_R^{t_2} \in \mathcal{W_R}. \\
\end{split}
\end{equation}
It shows that $\boldsymbol{W}_R^{t_2}$ and $\boldsymbol{W}_D^{t_2}$ are decoupled. Given $\boldsymbol{W}_D^{t_2}$, the solution of $\boldsymbol{W}_R^{t_2}$ can be expressed as
\begin{equation}\label{WRt2}
\boldsymbol{W}_R^{t_2} = \textrm{arg} ({\boldsymbol{W}_D^{t_2}}^H \widetilde{\boldsymbol{W}}^{t_2} , \mathcal{W_R})
\end{equation}
where $\textrm{arg}(\boldsymbol{A} , \mathcal{R})$ first normalizes each entry of $\boldsymbol{A}$ and then quantizes the normalized matrix in terms of $\mathcal{R}$. Note that the quantization is required since the resolution of phase shifters is limited in practice, e.g., the resolution is 2/64 if the phase shifter is 6 bits and the range of the angle is [-1,1].

Similarly, given $\boldsymbol{W}_R^{t_2}$, the optimization of $\boldsymbol{W}_D^{t_2}$ based on (\ref{minW32}) can be expressed as
\begin{equation}\label{OPP}
\begin{split}
\underset{ \boldsymbol{W}_D^{t_2}} {\bf min}  \quad & \| \boldsymbol{W}_R^{t_2} - {\boldsymbol{W}_D^{t_2}}^H \widetilde{\boldsymbol{W}}^{t_2} \| _F^2 \\
{\bf s.t.} \quad & {\boldsymbol{W}_D^{t_2}}^H \boldsymbol{W}_D^{t_2} = \beta^{-1} \boldsymbol{I}_{N_R} . \\
\end{split}
\end{equation}
It is seen that (\ref{OPP}) is similar to the orthogonal Procrustes problem \cite{gower2004procrustes}. Then the solution to (\ref{OPP}) can be obtained as
\begin{equation}\label{Alg1WBt2A}
\boldsymbol{W}_D^{t_2} = \beta^{-1/2}\boldsymbol{VU}^H,
\end{equation}
where $\boldsymbol{W}_R^{t_2} (\widetilde{\boldsymbol{W}}^{t_2})^H = \boldsymbol{U\Sigma V}^H$ represents the SVD of $\boldsymbol{W}_R^{t_2} (\widetilde{\boldsymbol{W}}^{t_2})^H$. Then we obtain $\boldsymbol{W}_B^{t_2}=\beta\boldsymbol{W}_D^{t_2}$. Note that both (\ref{WRt2}) and (\ref{Alg1WBt2A}) are optimal closed-form solutions to the problems expressed in (\ref{minW32}) and (\ref{OPP}), respectively.

As shown in $\mathbf{Algorithm~\ref{alg1}}$, we propose an algorithm of hybrid combining matrix design for the IA-based channel estimation. We repeatedly fix $\boldsymbol{W}_R^{t_2}$ to obtain $\boldsymbol{W}_B^{t_2}$ via (\ref{Alg1WBt2A}), and then fix $\boldsymbol{W}_R^{t_2}$ to obtain $\boldsymbol{W}_B^{t_2}$ via (\ref{WRt2}) in turn. Define the normalized iteration error $\epsilon$ as
\begin{equation}\label{epsilon}
\epsilon \triangleq \frac{ \| \boldsymbol{W}_R^{t_2,i} - \boldsymbol{W}_R^{t_2,i-1} \|_F^2 + \| \boldsymbol{W}_B^{t_2,i} - \boldsymbol{W}_B^{t_2,i-1} \|_F^2}{ \|\boldsymbol{W}_R^{t_2,i-1} \|_F^2 + \|\boldsymbol{W}_B^{t_2,i-1} \|_F^2},
\end{equation}
the stop condition is that the iterative update of both $\boldsymbol{W}_R^{t_2}$ and $\boldsymbol{W}_B^{t_2}$ is stable, i.e., $\epsilon < \delta$, where $\delta$ is the threshold.

\begin{algorithm}[!t]
	\caption{Hybrid Combining Matrix Design for IA-based Channel Estimation}
	\label{alg1}
	\begin{algorithmic}[1]
		\STATE \emph{Input:} $\boldsymbol{D}(N_A,K)$, $\mathcal{W_R}$, $P_W$, $\delta$, $\widetilde{\boldsymbol{W}}$.

        \STATE Obtain $\widetilde{\boldsymbol{W}}^{t_2}$ based on $\widetilde{\boldsymbol{W}}$ via (\ref{DefinitionW}), $t_2=1,2,\ldots,T_2$.
        \FOR{$t_2=1,2,\ldots,T_2$}
            \STATE Set $i \leftarrow 0$, and obtain $\boldsymbol{W}_R^{t_2,i}$ randomly from $\mathcal{W_R}$.
            \REPEAT
                \STATE $i \leftarrow i+1$.
                \STATE Fix $\boldsymbol{W}_R^{t_2,i-1}$ and obtain $\boldsymbol{W}_B^{t_2,i}$ via (\ref{Alg1WBt2A}).
                \STATE Fix $\boldsymbol{W}_B^{t_2,i}$ and obtain $\boldsymbol{W}_R^{t_2,i}$ via (\ref{WRt2}).
            \UNTIL{$\epsilon < \delta$}
            \STATE Update $\boldsymbol{W}_B^{t_2}$ via (\ref{Alg1SecondConstraintAdjustment}).
            \STATE Obtain $\boldsymbol{W}^{t_2}$ via (\ref{Wt2}).
        \ENDFOR
        \STATE Obtain $\boldsymbol{W}$ based on $\boldsymbol{W}^{t_2},~t_2=1,2,\ldots,T_2$ via (\ref{DefinitionW}).

        \STATE \emph{Output:} $\boldsymbol{W}$.
	\end{algorithmic}
\end{algorithm}

The optimization problem for \textbf{hybrid precoding matrix design} is
\begin{equation}\label{minF}
\begin{split}
\underset{ \boldsymbol{F}_u} {\bf min}  \quad & \| \boldsymbol{F}_u \boldsymbol{F}_u^H \boldsymbol{D}(M_A,K) - \gamma_M \boldsymbol{D}(M_A,K) \| _F \\
{\bf s.t.} \quad & \boldsymbol{F}_{R,u}^{t_1} \in \mathcal{F_R},~t_1 = 1,2,\ldots,T_1, \\
                 & \| \boldsymbol{F}_{R,u}^{t_1} \boldsymbol{F}_{B,u}^{t_1} \| _F ^2= P_F, ~t_1 = 1,2,\ldots,T_1, \\
\end{split}
\end{equation}
where $\mathcal{F_R}$ is the set of all feasible analog precoding matrix and $P_F$ is the given power for the hybrid precoding. Similar to (\ref{minW2}), (\ref{minF}) can be rewritten as
\begin{equation}\label{minF2}
\begin{split}
\underset{ \boldsymbol{F}_u} {\bf min}  \quad & \| \boldsymbol{F}_u \boldsymbol{F}_u^H  - \gamma_M \boldsymbol{I}_{M_A} \| _F \\
{\bf s.t.} \quad & \boldsymbol{F}_{R,u}^{t_1} \in \mathcal{F_R},~t_1 = 1,2,\ldots,T_1, \\
                 & \| \boldsymbol{F}_{R,u}^{t_1} \boldsymbol{F}_{B,u}^{t_1} \| _F ^2= P_F, ~t_1 = 1,2,\ldots,T_1. \\
\end{split}
\end{equation}
We can obtain a solution as $\widetilde{\boldsymbol{F}}_u$ by selecting the first $T_1$ columns of $\sqrt{\gamma_M}\boldsymbol{U}$, where $\boldsymbol{U}$ is any $M_A \times M_A$ unitary matrix. According to (\ref{DefinitionF}), we can obtain $\widetilde{\boldsymbol{f}}^{t_1}_u$ as the $t_1$-th column of $\widetilde{\boldsymbol{F}}_u$,~$t_1=1,2,\ldots,T_1$. Define
\begin{equation}\label{fBut1FBut1}
  \boldsymbol{f}_{B,u}^{t_1} \triangleq \boldsymbol{F}_{B,u}^{t_1} \textbf{1}_{M_R},~t_1=1,2,\ldots,T_1.
\end{equation}
Then we have
\begin{equation}
  \boldsymbol{f}^{t_1}_u  = \boldsymbol{F}_{R,u}^{t_1} \boldsymbol{f}_{B,u}^{t_1}.
\end{equation}
Similar to (\ref{minW3}), (\ref{minF2}) can be converted to $T_1$ subproblems, where each subproblem is expressed as
\begin{equation}\label{minF3}
\begin{split}
\underset{ \boldsymbol{F}_{R,u}^{t_1},~\boldsymbol{f}_{B,u}^{t_1} } {\bf min}  \quad & \| \boldsymbol{F}_{R,u}^{t_1} \boldsymbol{f}_{B,u}^{t_1}  - \widetilde{\boldsymbol{f}}^{t_1}_u \| _F \\
{\bf s.t.} \quad & \boldsymbol{F}_{R,u}^{t_1} \in \mathcal{F_R}, ~\| \boldsymbol{F}_{R,u}^{t_1} \boldsymbol{F}_{B,u}^{t_1} \| _F ^2= P_F. \\
\end{split}
\end{equation}
Similar to (\ref{minW3}), the second constraint of (\ref{minF3}) can also be temporarily neglected. Therefore, we may replace $\boldsymbol{W}_B^{t_2}$, $\boldsymbol{W}_R^{t_2}$, $\widetilde{\boldsymbol{W}}^{t_2}$, $\mathcal{W_R}$ and $P_W$ in (\ref{minW3}) with $(\boldsymbol{f}_{B,u}^{t_1})^H$, $(\boldsymbol{F}_{R,u}^{t_1})^H$, $(\widetilde{\boldsymbol{f}}^{t_1}_u)^H$, $\mathcal{F_R}$ and $P_F$, respectively.

In order to run $\mathbf{Algorithm~\ref{alg1}}$ to obtain $\boldsymbol{F}_u$, we have to further replace $N_A$, $T_2$,
$\widetilde{\boldsymbol{W}}$ and $\boldsymbol{W}$ with $M_A$, $T_1$, $\widetilde{\boldsymbol{F}}_u$ and $\boldsymbol{F}_u$. The routine of the algorithm is exactly the same, except that an additional operation to obtain $\boldsymbol{F}_{B,u}^{t_1}$ as
\begin{equation}
  \boldsymbol{F}_{B,u}^{t_1} =  \boldsymbol{f}_{B,u}^{t_1} \textbf{1}_{M_R}^T / M_R
\end{equation}
is required after finishing step~9; and $\boldsymbol{W}_B^{t_2}$ should be replaced by $(\boldsymbol{F}_{B,u}^{t_1})^H$ at step~10.

In summary, in the first half of the IA-based channel estimation, we design of hybrid combining matrix and hybrid precoding matrix; while in the other half to be discussed, we will search the largest entry of the over-sampled beamspace channel matrix.

\begin{figure}[!t]
\centering
\includegraphics[width=75mm]{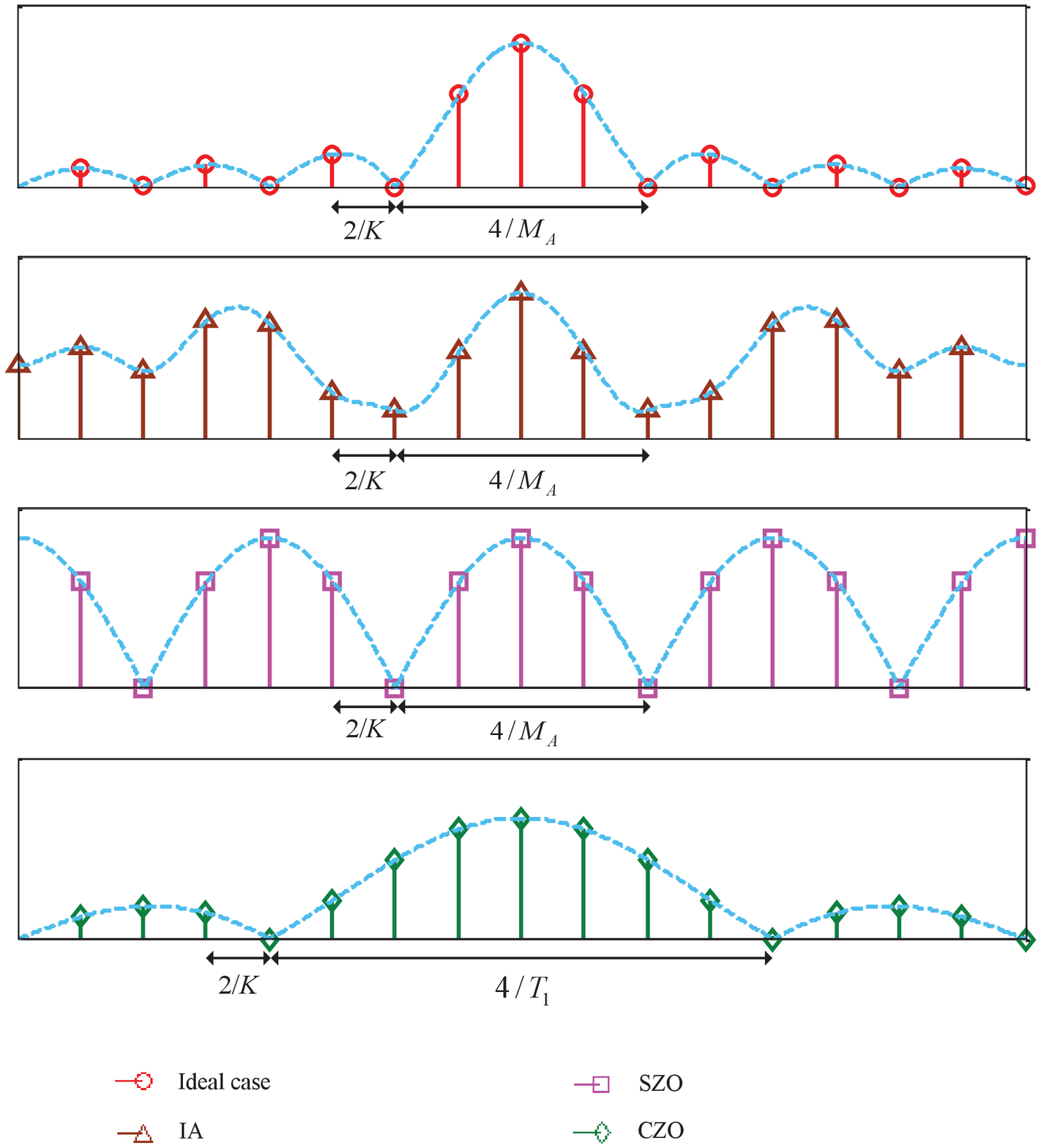}
\caption{Amplitude of $\boldsymbol{r}_M$.}
\label{FIG_1}
\end{figure}

\subsubsection{Search the Largest Entry}
Instead of exhaustively search the largest entry from $\boldsymbol{R}_u^v$ defined in (\ref{Ruv}), we can improve the efficiency of the search algorithm by analyzing the structure of $\boldsymbol{R}_u^v$. Neglecting the term from the additive noise and assuming there is single path, we rewrite $\boldsymbol{R}_u^v$ as
\begin{equation}
\begin{split}
   \boldsymbol{R}_u^v &= \boldsymbol{D}(N_A,K)^H \boldsymbol{W}^H \boldsymbol{W} \boldsymbol{H}_u \boldsymbol{F}_u \boldsymbol{F}_u^H \boldsymbol{D}(M_A,K)\\
     &= \sqrt{N_AM_A}g_{u,i} \boldsymbol{r}_N \boldsymbol{r}_M^H\\
\end{split}
\end{equation}
where
\begin{equation}\label{rN}
\boldsymbol{r}_N \triangleq \boldsymbol{D}(N_A,K)^H \boldsymbol{W}^H \boldsymbol{W} \boldsymbol{\alpha} (N_A,\theta_{u,1}),
\end{equation}
\begin{equation}\label{rM}
\boldsymbol{r}_M^H \triangleq \boldsymbol{\alpha}^{H} (M_A,\varphi_{u,1}) \boldsymbol{F}_u \boldsymbol{F}_u^H \boldsymbol{D}(M_A,K).
\end{equation}
Since $\boldsymbol{r}_N$ is a column vector and $\boldsymbol{r}_M^H$ is a row vector, the largest entry of $\boldsymbol{R}_u^v$ essentially depends on the largest entry of $\boldsymbol{r}_N$ and $\boldsymbol{r}_M$. Therefore we will analyze the structure of $\boldsymbol{r}_N$ and $\boldsymbol{r}_M$.

In the ideal case, i.e., $\boldsymbol{F}_u \boldsymbol{F}_u^H \boldsymbol{D}(M_A,K) = \gamma_M \boldsymbol{D}(M_A,K)$, $\boldsymbol{r}_M$ is an over-sampled transmit steering vector of $\boldsymbol{\alpha} (M_A,\varphi_{u,1})$ with an interval of $2/K$. As shown in Fig.~\ref{FIG_1}, we illustrate the envelope of $\boldsymbol{r}_M$, where the position corresponding to the peak of $\boldsymbol{r}_M$ is the channel AoD with quantization error of $2/K$. In order to apply fast algorithms such as trichotomy search to find the peak of the curve, we have to first find the main lobe with the width of $4/M_A$; otherwise these algorithms may stop the search at the peak of side lobes.

Now we propose $\mathbf{Algorithm~\ref{alg2}}$ to fast search the largest entry of $\boldsymbol{R}_u^v$, considering the structure of the steering vectors. $\mathbf{Algorithm~\ref{alg2}}$ includes two stages. In the \textbf{first stage}, we find the main lobe of $\boldsymbol{r}_M$ and $\boldsymbol{r}_N$ without oversampling. In the \textbf{second stage} with oversampling, we apply the trichotomy search to find the peak of the main lobe.

\begin{algorithm}[!t]
	\caption{Searching the Largest Entry Corresponding to the AoA and AoD of LOS Path}
	\label{alg2}
	\begin{algorithmic}[1]
		\STATE \emph{Input:} $\boldsymbol{R}_u$.
        \STATE \textbf{(First Stage)}
        \STATE Obtain $\bar{\boldsymbol{R}}_u^v$ via (\ref{bar_Ruv}).
        \STATE Obtain $s_{q}$ and $s_{p}$ via (\ref{find_sq}) and (\ref{find_sp}), respectively.
        \STATE Obtain $\boldsymbol{\Gamma} = [\Gamma_1,\Gamma_2]$ and $\boldsymbol{\Upsilon} = [\Upsilon_1,\Upsilon_2]$ via (\ref{Gamma}).
        \STATE \textbf{(Second Stage)}
        \STATE Obtain $\boldsymbol{R}_u^v$ via (\ref{Ruv}).
        \WHILE{$\Gamma_2 - \Gamma_1 > 2/K$ or $\Upsilon_2 - \Upsilon_1 > 2/K$}
            \STATE Obtain $Q_1$, $Q_2$, $Q_3$ and $Q_4$ via (\ref{Q14}).
            \STATE Obtain $Q_{min}$ via (\ref{Qmin}).
            \STATE Update $\boldsymbol{\Gamma}$ and $\boldsymbol{\Upsilon}$.
        \ENDWHILE
        \STATE $\hat{\theta}_{u,1}=\Gamma_1$, $\hat{\varphi}_{u,1}=\Upsilon_1$
        \STATE \emph{Output:} $\hat{\theta}_{u,1}$, $\hat{\varphi}_{u,1}$.
	\end{algorithmic}
\end{algorithm}

\begin{figure}[!t]
\centering
\includegraphics[width=90mm]{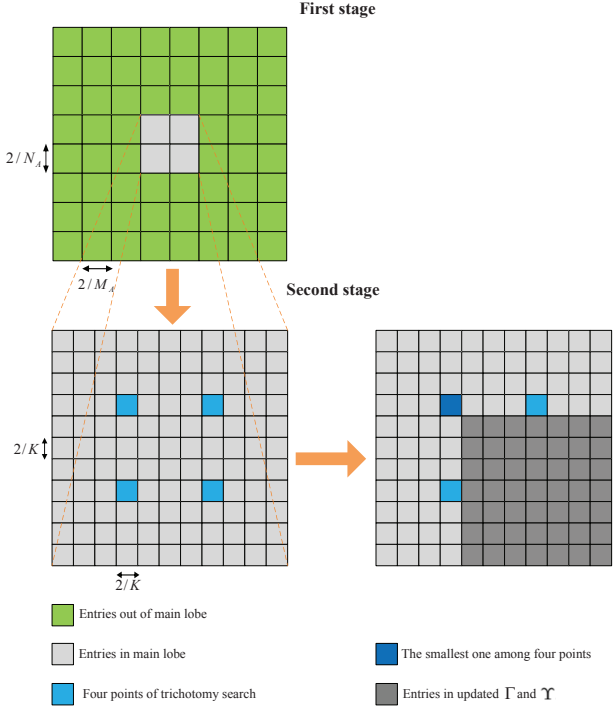}
\caption{Illustration of $\mathbf{Algorithm~\ref{alg2}}$.}
\label{FIG_2}
\end{figure}

In the \textbf{first stage} from step~3 to step~6, we search the main lobe, which is formed by two adjacent columns and two adjacent rows of $\bar{\boldsymbol{H}}_u^v$ defined in (\ref{bar_Huv}) \cite{ma2017channel}. We obtain the beamspace receiving matrix $\bar{\boldsymbol{R}}_u^v \in{\mathbb{C}^{N_A \times{M_A}}}$ at step~3 as
\begin{equation}\label{bar_Ruv}
\bar{\boldsymbol{R}}_u^v = \boldsymbol{D}(N_A,N_A)^H \boldsymbol{W}^H \boldsymbol{W} \boldsymbol{H}_u \boldsymbol{F}_u \boldsymbol{F}_u^H \boldsymbol{D}(M_A,M_A) + \bar{\boldsymbol{n}}^v
\end{equation}
where $\bar{\boldsymbol{n}}^v \triangleq \boldsymbol{D}(N_A,N_A)^H \boldsymbol{W}^H \widetilde{\boldsymbol{n}} \boldsymbol{F}_u^H \boldsymbol{D}(M_A,M_A)$ is a noise term. We first find two adjacent columns indexed by \{$s_q$,~$s_q+1$\} with the largest channel power from $\bar{\boldsymbol{H}}_u^v$ at step~4 via
\begin{equation}\label{find_sq}
s_{q} = \arg\underset{q=1,2,\ldots,M_A-1} {\max} \| \bar{\boldsymbol{R}}_{u,q}^v \|_F
\end{equation}
where $\bar{\boldsymbol{R}}_{u,q}^v \in{\mathbb{C}^{N_A\times{2}}}$ represents a submatrix consisted of two consecutive columns of $\bar{\boldsymbol{R}}_u^v$, with column indices denoted as $q$ and $q+1$. Similarly, we find two adjacent rows indexed by \{$s_p$,$s_p+1$\} with the largest channel power from $\bar{\boldsymbol{R}}_u^v$ at step~4 via
\begin{equation}\label{find_sp}
s_{p} = \arg\underset{p=1,2,\ldots,N_A-1} {\max} \| \bar{\boldsymbol{R}}_{u,p}^v \|_F
\end{equation}
where $\bar{\boldsymbol{R}}_{u,p}^v \in{\mathbb{C}^{2\times{M_A}}}$ represents a submatrix consisted of two consecutive rows of $\bar{\boldsymbol{R}}_u^v$, with row indices denoted as $p$ and $p+1$. By finding out the largest two adjacent columns \{$s_q$,~$s_q+1$\} and rows \{$s_p$,~$s_p+1$\} of beamspace channel matrix $\bar{\boldsymbol{H}}_u^v$, the search of AoA and AoD can be limited to the range of
\begin{equation}\label{Gamma}
\boldsymbol{\Gamma} = [\Gamma_1,\Gamma_2],~\boldsymbol{\Upsilon} = [\Upsilon_1,\Upsilon_2],
\end{equation}
respectively, where
\begin{equation}
  \begin{split}
     \Gamma_1 \triangleq & -1+2(s_p-3/2)/N_A, \Gamma_2 \triangleq  -1+2(s_p+1/2)/N_A, \\
     \Upsilon_1 \triangleq & -1+2(s_q-3/2)/M_A, \Upsilon_2 \triangleq   -1+2(s_q+1/2)/M_A.
  \end{split}
\end{equation}
In this way, we narrow down the search space of the AoA and AoD from $[-1,1]$ to $\boldsymbol{\Gamma}$ and $\boldsymbol{\Upsilon}$, respectively. As shown in Fig.~\ref{FIG_2}, the two adjacent rows and two adjacent columns found in the first stage are illustrated in a grey square area.

In the \textbf{second stage} from step~8 to step~14, we find the coordinates of the largest entry of $\boldsymbol{R}_u^v$ corresponding to the AoA in $\boldsymbol{\Gamma}$  and AoD in $\boldsymbol{\Upsilon}$. Note that the quantization error of the AoD and AoA is reduced from $2/M_A$ and $2/N_A$ to both $2/K$ by oversampling. We apply the trichotomy search. The two points that divide $\boldsymbol{\Gamma}$ into three equal parts are $2\Gamma_1/3 + \Gamma_2/3$ and $\Gamma_1/3 + 2\Gamma_2/3$. Similarly, the two points that divide $\boldsymbol{\Upsilon}$ into three equal parts are $2\Upsilon_1/3 + \Upsilon_2/3$ and $\Upsilon_1/3 + 2\Upsilon_2/3$. These four points marked in light blue in Fig.~\ref{FIG_2} can divide the area of $\boldsymbol{R}_u^v$ corresponding to $\boldsymbol{\Gamma}$ and $\boldsymbol{\Upsilon}$ into nine smaller areas. The entries corresponding to these four points are
\begin{align}\label{Q14}
& Q_1 = \boldsymbol{R}_u^v[ {quan}(2\Gamma_1/3 + \Gamma_2/3) , {quan}(2\Upsilon_1/3 + \Upsilon_2/3 )], \notag \\
& Q_2 = \boldsymbol{R}_u^v[ {quan}(2\Gamma_1/3 + \Gamma_2/3) , {quan}(\Upsilon_1/3 + 2\Upsilon_2/3 )], \notag \\
& Q_3 = \boldsymbol{R}_u^v[ {quan}(\Gamma_1/3 + 2\Gamma_2/3) , {quan}(2\Upsilon_1/3 + \Upsilon_2/3 )], \notag \\
& Q_4 = \boldsymbol{R}_u^v[ {quan}(\Gamma_1/3 + 2\Gamma_2/3) , {quan}(\Upsilon_1/3 + 2\Upsilon_2/3 )], \notag \\
\end{align}
where ${quan}()$ is the quantization function to quantize the consecutive $\theta$ into $K$ discrete points, which is defined as
\begin{equation}
{quan}(\theta) \triangleq \left\langle K(\theta+1) / 2 \right\rangle.
\end{equation}
Then we compare the amplitude of $Q_1$, $Q_2$, $Q_3$ and $Q_4$ to find the smallest one, which is expressed as
\begin{equation}\label{Qmin}
Q_{min} = \min \big\{|Q_1|, |Q_2|, |Q_3|, |Q_4|\big\}.
\end{equation}
We delete the parts that include $Q_{min}$ and update $\boldsymbol{\Gamma}$ and $\boldsymbol{\Upsilon}$. As shown in Fig.~\ref{FIG_2}, $Q_{min}$ is marked in dark blue, where the entries within the area of the updated $\boldsymbol{\Gamma}$ and $\boldsymbol{\Upsilon}$ are marked in dark grey. For example, if $Q_{min}=Q_1$, the updated $\Gamma_1$, $\Gamma_2$, $\Upsilon_1$ and $\Upsilon_2$, denoted as $\bar{\Gamma}_1$, $\bar{\Gamma}_2$, $\bar{\Upsilon}_1$ and $\bar{\Upsilon}_2$, respectively, can be represented as
\begin{equation}\label{updategam}
\begin{split}
& \bar{\Gamma}_1 = 2\Gamma_1/3 + \Gamma_2/3,~\bar{\Gamma}_2 = \Gamma_2, \\
& \bar{\Upsilon}_1 = 2\Upsilon_1/3 + \Upsilon_2/3,~\bar{\Upsilon}_2 = \Upsilon_2. \\
\end{split}
\end{equation}
We repeat the procedures until $\Gamma_2 - \Gamma_1 \leq 2/K$ and $\Upsilon_2 - \Upsilon_1 \leq 2/K$, which means the resolution $2/K$ of over-sampling search is reached for both the AoA and AoD. Finally we output the estimated AoA and AoD of LOS path at step~15.

\subsection{ZO-based Beamspace Channel Estimation Schemes}
In the IA-based beamspace channel estimation, we first solve two optimization problems (\ref{minW}) and (\ref{minF}) to design the hybrid combining matrix and hybrid precoding matrix. However, considering that $\boldsymbol{W}^H \boldsymbol{W}$ does not exactly equal $\gamma_N \boldsymbol{I}_{N_A}$ and $\boldsymbol{F}_u \boldsymbol{F}_u^H$ does not exactly equal $\gamma_M \boldsymbol{I}_{M_A}$, there are possible errors for the search of the largest entry even without noise. Also note that the off-diagonal entries of $\boldsymbol{W}^H \boldsymbol{W}$ and $\boldsymbol{F}_u \boldsymbol{F}_u^H$ are not restricted to be zero, which may introduce some interference for beamspace channel estimation.

In this subsection, we design hybrid precoding matrix $\boldsymbol{F}_u$ and hybrid combining matrix $\boldsymbol{W}$ so that the coordinates of the largest entry of $\boldsymbol{R}_u^v$ are the AoA and AoD of the LOS path with the quantization error of $2/K$ for single path. Since the largest entry of $\boldsymbol{R}_u^v$ essentially depends on the largest entry of $\boldsymbol{r}_N$ in (\ref{rN}) and $\boldsymbol{r}_M$ in (\ref{rM}), respectively, we will first analyze $\boldsymbol{r}_N$ and $\boldsymbol{r}_M$.

The $k$th entry of $\boldsymbol{r}_M$ can be represented as $\boldsymbol{r}_M[k] = \boldsymbol{\alpha}^H (M_A,\varphi_{u,1}) \boldsymbol{F}_u \boldsymbol{F}_u^H \boldsymbol{\alpha}(M_A,-1+2(k-1)/K)$. Therefore, $\boldsymbol{F}_u$ should be designed so that the largest entry of $\boldsymbol{r}_M$ is $\boldsymbol{r}_M[{quan}(\varphi_{u,1})]$, which indicates the quantization error is $2/K$. However, it is difficult to solve this problem for discrete $K$ points. Now we convert this discrete problem into continuous problem to obtain the derivative. To ease the notation, we define $\boldsymbol{\alpha}_M (\varphi) \triangleq \boldsymbol{\alpha} (M_A,\varphi)$ and $\varphi_g \triangleq \varphi_{u,1}$. We further define a function of $\varphi$ as $ \mathcal{R}(\varphi_g,\varphi) \triangleq \big| \boldsymbol{\alpha}_M^H (\varphi_g) \boldsymbol{F}_u \boldsymbol{F}_u^H \boldsymbol{\alpha}_M (\varphi) \big|^2$, which can be regarded as the continuous version of $\boldsymbol{r}_M$. The problem can be formulated as
\begin{equation}\label{ContinuousProblem}
  \underset{\varphi} {\bf arg max} ~ \mathcal{R} (\varphi_g,\varphi).
\end{equation}
where $\varphi_g$ is the genuine channel parameter defined in (\ref{ULAchannelmodel}) but is unknown to the receiver. It is expected that one of the solutions to (\ref{ContinuousProblem}) is $\varphi_g$.

Define $\boldsymbol{F}^{+}_u \triangleq \boldsymbol{F}_u \boldsymbol{F}_u^H$. Note that $\boldsymbol{F}^{+}_u$ is a Hermitian matrix, i.e., ${\boldsymbol{F}^{+}_u}^H = \boldsymbol{F}^{+}_u$. The partial derivative of $\mathcal{R}(\varphi_g,\varphi)$ over $\varphi$ is
\begin{align}\label{derivative}
  \frac{\partial{ \mathcal{R}(\varphi_g,\varphi) }}  {\partial{\varphi}}  &= \frac{\partial{ \boldsymbol{\alpha}^H_M (\varphi) \boldsymbol{F}^{+}_u \boldsymbol{\alpha}_M(\varphi_g) \boldsymbol{\alpha}^H_M (\varphi_g) \boldsymbol{F}^{+}_u \boldsymbol{\alpha}_M(\varphi) }}  {\partial{\varphi}} \notag \\
  &= \frac{\partial{ \boldsymbol{\alpha}^H_M (\varphi) }}  {\partial{\varphi}} \boldsymbol{F}^{+}_u \boldsymbol{\alpha}_M(\varphi_g) \boldsymbol{\alpha}^H_M (\varphi_g) \boldsymbol{F}^{+}_u \boldsymbol{\alpha}_M(\varphi) + \notag \\
  &~~~~ \boldsymbol{\alpha}^H_M (\varphi)  \boldsymbol{F}^{+}_u \boldsymbol{\alpha}_M(\varphi_g) \boldsymbol{\alpha}^H_M (\varphi_g) \boldsymbol{F}^{+}_u \frac{\partial{ \boldsymbol{\alpha}_M(\varphi) }}  {\partial{\varphi}}.
\end{align}
Define a diagonal matrix $\boldsymbol{B}\in{\mathbb{C}^{M_A\times{M_A}}}$ as
\begin{equation}
  \boldsymbol{B} \triangleq
  \begin{bmatrix}
    0 & 0     & \cdots & 0 \\
    0 & -j\pi & \cdots &  0       \\
    \vdots & \vdots & \ddots & \vdots\\
    0 & 0 & \cdots & -j\pi(M_A-1)
  \end{bmatrix}
\end{equation}
which is a diagonal matrix with $M_A-1$ nonzero entries. Then (\ref{derivative}) can be further written as
\begin{align}\label{derivative2}
   \frac{\partial{ \mathcal{R}(\varphi_g,\varphi) }}  {\partial{\varphi}}
  &= \boldsymbol{\alpha}^H_M(\varphi) \boldsymbol{B}^H \boldsymbol{F}^{+}_u \boldsymbol{\alpha}_M(\varphi_g) \boldsymbol{\alpha}^H_M (\varphi_g) \boldsymbol{F}^{+}_u \boldsymbol{\alpha}_M(\varphi) + \notag \\
  &~~~~ \boldsymbol{\alpha}^H_M (\varphi)  \boldsymbol{F}^{+}_u \boldsymbol{\alpha}_M(\varphi_g) \boldsymbol{\alpha}^H_M (\varphi_g) \boldsymbol{F}^{+}_u \boldsymbol{B}\boldsymbol{\alpha}_M(\varphi) \notag \\
  &= 2 \mathbb{Re}\left\{ \boldsymbol{\alpha}^H_M (\varphi)  \boldsymbol{F}^{+}_u \boldsymbol{\alpha}_M(\varphi_g) \boldsymbol{\alpha}^H_M (\varphi_g) \boldsymbol{F}^{+}_u \boldsymbol{B}\boldsymbol{\alpha}_M(\varphi) \right\}.
\end{align}
Since $\varphi_g$ is one of the solutions to (\ref{ContinuousProblem}), we have
\begin{equation}\label{DerivativeEqualsZero}
  \left. \frac{\partial{ ~\mathcal{R}(\varphi_g,\varphi) }}  {\partial{\varphi}} \right|_{\varphi = \varphi_g} = 0.
\end{equation}
Then we have
\begin{equation}
   2\mathbb{Re}\left\{ \boldsymbol{\alpha}^H_M (\varphi_g)  \boldsymbol{F}^{+}_u \boldsymbol{\alpha}_M(\varphi_g) \boldsymbol{\alpha}^H_M (\varphi_g) \boldsymbol{F}^{+}_u \boldsymbol{B}\boldsymbol{\alpha}_M(\varphi_g) \right\}=0.
\end{equation}
Since $\boldsymbol{F}^{+}_u$ is a Hermitian matrix, $\boldsymbol{\alpha}^H_M (\varphi_g)  \boldsymbol{F}^{+}_u \boldsymbol{\alpha}_M(\varphi_g)$ is a real number. Therefore we have
\begin{equation}\label{Realpha}
    \boldsymbol{\alpha}^H_M (\varphi_g)  \boldsymbol{F}^{+}_u \boldsymbol{\alpha}_M(\varphi_g) 2\mathbb{Re} \left\{\boldsymbol{\alpha}^H_M (\varphi_g) \boldsymbol{F}^{+}_u \boldsymbol{B}\boldsymbol{\alpha}_M(\varphi_g) \right\}=0.
\end{equation}
There are two solutions of $\boldsymbol{F}^{+}_u$ to (\ref{Realpha}), denoted as \textbf{Solution 1} and \textbf{Solution 2}. \textbf{Solution 1} of $\boldsymbol{F}^{+}_u$ satisfies
\begin{equation}\label{solution1}
  \boldsymbol{\alpha}^H_M (\varphi_g)  \boldsymbol{F}^{+}_u \boldsymbol{\alpha}_M(\varphi_g)=0,
\end{equation}
while \textbf{Solution 2} of $\boldsymbol{F}^{+}_u$ satisfies
\begin{equation}\label{solution2}
  2\mathbb{Re} \left\{\boldsymbol{\alpha}^H_M (\varphi_g) \boldsymbol{F}^{+}_u \boldsymbol{B}\boldsymbol{\alpha}_M(\varphi_g) \right\}=0.
\end{equation}

In terms of \textbf{Solution 1}, if (\ref{solution1}) is satisfied, we set $\varphi_g=-1,-1+2/M_A,\ldots,-1+2(M_A-1)/M_A$ and obtain $M_A$ equations. We combine these $M_A$ equations together, having
\begin{equation}\label{solution11}
  \boldsymbol{D}(M_A,M_A)^H  \boldsymbol{F}^{+}_u \boldsymbol{D}(M_A,M_A) = \boldsymbol{A}_1
\end{equation}
where $\boldsymbol{A}_1\in{\mathbb{C}^{M_A \times{M_A}}}$ is a matrix with zero diagonal entries, leading to
\begin{equation}\label{TrA}
  \textrm{Tr}(\boldsymbol{A}_1)=0.
\end{equation}
But the trace of the expression on the left side of (\ref{solution11}) is
\begin{align}\label{TrF}
  & \textrm{Tr}\big(\boldsymbol{D}(M_A,M_A)^H  \boldsymbol{F}^{+}_u \boldsymbol{D}(M_A,M_A)\big)  \notag \\
  &= \textrm{Tr}\big(\boldsymbol{D}(M_A,M_A)\boldsymbol{D}(M_A,M_A)^H  \boldsymbol{F}^{+}_u \big) \notag \\
  &= \textrm{Tr}\big( \boldsymbol{F}^{+}_u \big) / M_A \notag \\
  &\overset{(a)}{\geq}0
\end{align}
where $(a)$ is true because $\boldsymbol{F}^{+}_u$ is positive semi-definite, and the equality of $(a)$ holds only when $\boldsymbol{F}^{+}_u=\boldsymbol{0}_{M_A}$. Simultaneously satisfying (\ref{TrA}) and (\ref{TrF}) leads to
\begin{equation}
  \boldsymbol{F}^{+}_u=\boldsymbol{0}_{M_A}.
\end{equation}
However, in practice $\boldsymbol{F}^{+}_u$ can not be zero matrix due to the power constraint in (\ref{minF}). Therefore \textbf{Solution 1} is meaningless.

In terms of \textbf{Solution 2}, we first rewrite the expression on the left side of (\ref{solution2}) as
\begin{align}
  & 2\mathbb{Re} \left\{\boldsymbol{\alpha}^H_M (\varphi_g) \boldsymbol{F}^{+}_u \boldsymbol{B}\boldsymbol{\alpha}_M(\varphi_g) \right\} \notag \\
  &= \boldsymbol{\alpha}^H_M (\varphi_g) \boldsymbol{B}^H \boldsymbol{F}^{+}_u \boldsymbol{\alpha}_M(\varphi_g) + \boldsymbol{\alpha}^H_M (\varphi_g) \boldsymbol{F}^{+}_u \boldsymbol{B} \boldsymbol{\alpha}_M(\varphi_g) \notag \\
  &= \frac{\partial{ \boldsymbol{\alpha}^H_M (\varphi_g) }}  {\partial{\varphi_g}} \boldsymbol{F}^{+}_u \boldsymbol{\alpha}_M(\varphi_g) + \boldsymbol{\alpha}^H_M (\varphi_g) \boldsymbol{F}^{+}_u \frac{\partial{ \boldsymbol{\alpha}_M(\varphi_g) }}  {\partial{\varphi_g}} \notag \\
  &= \frac{\partial{ \boldsymbol{\alpha}^H_M (\varphi_g) \boldsymbol{F}^{+}_u \boldsymbol{\alpha}_M(\varphi_g)  }}  {\partial{\varphi_g}}.
\end{align}
Therefore (\ref{solution2}) can be further written as
\begin{equation}\label{alphaF}
  \boldsymbol{\alpha}^H_M (\varphi_g) \boldsymbol{F}^{+}_u \boldsymbol{\alpha}_M(\varphi_g) = \xi.
\end{equation}
where $\xi$ is a constant. Similar to (\ref{solution11}), we have
\begin{equation}\label{solution22}
  \boldsymbol{D}(M_A,M_A)^H  \boldsymbol{F}^{+}_u \boldsymbol{D}(M_A,M_A) = \boldsymbol{A}_2
\end{equation}
where $\boldsymbol{A}_2\in{\mathbb{C}^{M_A \times{M_A}}}$ is a matrix whose diagonal entries all equal $\xi$, leading to
\begin{equation}\label{TrA2}
  \textrm{Tr}(\boldsymbol{A}_2)=\xi M_A.
\end{equation}
The trace of the expression on the left of (\ref{solution22}) is $\textrm{Tr}\big( \boldsymbol{F}^{+}_u \big) / M_A$ according to (\ref{TrF}). Combining (\ref{TrA2}) and (\ref{TrF}), we have
\begin{equation}\label{ValueOfXi}
  \xi = \textrm{Tr}\big( \boldsymbol{F}^{+}_u \big) / M_A^2.
\end{equation}
Now we solve (\ref{alphaF}) to obtain the solution of $\boldsymbol{F}^{+}$. The expression on the left side of (\ref{alphaF}) is essentially in a quadratic form, which can be written in detail as
\begin{align}\label{leftalphaF}
  & \boldsymbol{\alpha}^H_M (\varphi_g) \boldsymbol{F}^{+}_u \boldsymbol{\alpha}_M(\varphi_g)  \\ \notag
  &= \sum_{i=1}^{M_A} \sum_{l=1}^{M_A} \boldsymbol{\alpha}^H_M (\varphi_g)[i] \boldsymbol{F}^{+}_u[i,l] \boldsymbol{\alpha}_M(\varphi_g)[l]  \\ \notag
  &= \frac{1}{M_A^2} \sum_{i=1}^{M_A} \sum_{l=1}^{M_A} e^{-j\pi\varphi_g(l-i)} \boldsymbol{F}^{+}_u[i,l]  \\ \notag
  &= \frac{1}{M_A^2} \sum_{k=1-M_A }^{-1}  \left( \sum_{l=1}^{M_A+k} \boldsymbol{F}^{+}_u[l-k,l] \right) e^{-j\pi\varphi_gk} +  \\ \notag
  &~~~~ \frac{1}{M_A^2} \sum_{k=1 }^{M_A-1}  \left( \sum_{i=1}^{M_A-k} \boldsymbol{F}^{+}_u[i,i+k] \right) e^{-j\pi\varphi_gk} + \frac{\textrm{Tr}\big( \boldsymbol{F}^{+}_u \big)}{M_A^2}.
\end{align}
Based on (\ref{alphaF}), (\ref{ValueOfXi}) and (\ref{leftalphaF}), we have
\begin{align}
  & \frac{1}{M_A^2} \sum_{k=1-M_A }^{-1}  \left( \sum_{l=1}^{M_A+k} \boldsymbol{F}^{+}_u[l-k,l] \right) e^{-j\pi\varphi_gk} + \notag \\
  & \frac{1}{M_A^2} \sum_{k=1 }^{M_A-1}  \left( \sum_{i=1}^{M_A-k} \boldsymbol{F}^{+}_u[i,i+k] \right) e^{-j\pi\varphi_gk}=0
\end{align}
Since in practice $\varphi_g$ can be any value in [-1,1], $\boldsymbol{F}^{+}_u$ should satisfy
\begin{align}\label{Fuplus}
 & \sum_{l=1}^{M_A+k} \boldsymbol{F}^{+}_u[l-k,l] = 0,~k = 1-M_A,2-M_A,\cdots,-1, \notag \\
 & \sum_{i=1}^{M_A-k} \boldsymbol{F}^{+}_u[i,i+k] = 0,~k = 1,2,\cdots,M_A-1,
\end{align}
which means the summation of off-diagonal entries on the same line parallel to the main diagonal line of $\boldsymbol{F}^{+}_u$ is zero. To simplify the problem, we set all off-diagonal entries of $\boldsymbol{F}^{+}_u$ to be zero, a.k.a., zero off-diagonal (ZO). In this way, we have to shut down some antenna ports for ZO-based schemes, while the number of active RF chains keeps the same. In fact, according to the existing literature~\cite{rial2016hybird,zhai2017joint,xiao2016hi}, some antenna ports are shut down in mmWave massive MIMO systems. In~\cite{rial2016hybird} and~\cite{zhai2017joint}, some antenna ports are shut down to achieve the antenna selection aiming at the sum-rate maximization. In~\cite{xiao2016hi}, some antenna ports are shut down to achieve wide mainlobe of beams for mmWave beam training based on hierarchical codebooks. Since $\boldsymbol{F}_u \in{\mathbb{C}^{M_A \times T_1}}$ is a tall matrix with $M_A>T_1$ and $\boldsymbol{F}^{+}_u = \boldsymbol{F}_u \boldsymbol{F}_u^H$, the rank of $\boldsymbol{F}^{+}_u$ is no more than $T_1$, indicating that there are only $T_1$ nonzero entries of the diagonal matrix $\boldsymbol{F}^{+}_u$. We set the $T_1$ nonzero entries to be the same $\gamma_M\sqrt{M_A/T_1}$ and the left $M_A-T_1$ diagonal entries to be zero.

Given $\boldsymbol{F}^{+}_u$, we can obtain $\boldsymbol{F}_u$ by making SVD of $\boldsymbol{F}^{+}_u$, i.e., $\boldsymbol{F}^{+}_u=\boldsymbol{U}_M\boldsymbol{\Sigma}_M \boldsymbol{U}_M^H$, where $\boldsymbol{U}_M\in{\mathbb{C}^{M_A\times{T_1}}}$ is an unitary matrix and $\boldsymbol{\Sigma}_M\in{\mathbb{C}^{T_1\times{T_1}}}$ is a real diagonal matrix. Then we can obtain $\boldsymbol{F}_u$ by
\begin{equation}\label{obtainFu}
  \boldsymbol{F}_u =  \boldsymbol{U}_M\sqrt{\boldsymbol{\Sigma}_M}.
\end{equation}

Similar to $\boldsymbol{F}^{+}$, we define $\boldsymbol{W}^{+} \triangleq \boldsymbol{W}^H \boldsymbol{W}$, where $\boldsymbol{W}^{+}$ should satisfy
\begin{align}
 & \sum_{l=1}^{N_A+k} \boldsymbol{W}^{+}[l-k,l] = 0,~k = 1-N_A,2-N_A,\cdots,-1, \notag \\
 & \sum_{i=1}^{N_A-k} \boldsymbol{W}^{+}[i,i+k] = 0,~k = 1,2,\cdots,N_A-1.
\end{align}

We set $\boldsymbol{W}^{+}$ to be a diagonal matrix, where $T_3$ nonzero diagonal entries are set to be the same $\gamma_N\sqrt{N_A/T_3}$ and the left $N_A-T_3$ diagonal entries are zero. Given $\boldsymbol{W}^{+}$, we can obtain $\boldsymbol{W}$ by making SVD of $\boldsymbol{W}^{+}$, i.e., $\boldsymbol{W}^{+}=\boldsymbol{U}_N\boldsymbol{\Sigma}_N \boldsymbol{U}_N^H$, where $\boldsymbol{U}_N\in{\mathbb{C}^{N_A\times{T_3}}}$ is an unitary matrix and $\boldsymbol{\Sigma}_N\in{\mathbb{C}^{T_3\times{T_3}}}$ is a diagonal matrix. Then we can obtain $\boldsymbol{W}$ by
\begin{equation}\label{obtainW}
  \boldsymbol{W} =  \sqrt{\boldsymbol{\Sigma}_N} \boldsymbol{U}_N^H.
\end{equation}

According to different layout of nonzero diagonal entries, now we propose a scattered zero off-diagonal (SZO) and a concentrated zero off-diagonal (CZO) based beamspace channel estimation scheme. In the SZO scheme, the nonzero diagonal entries of $\boldsymbol{F}_u^{+}$ and $\boldsymbol{W}^{+}$ are uniformly distributed with the same interval. In the CZO scheme, the nonzero diagonal entries of $\boldsymbol{F}_u^{+}$ and $\boldsymbol{W}^{+}$ are concentrated on the upper left corner of the matrix.

\subsubsection{\bf SZO-based Beamspace Channel Estimation Scheme}
We design $\boldsymbol{F}^{+}_u$ as
\begin{equation}
  \boldsymbol{F}^{+}_u = \gamma_M\sqrt{\frac{M_A}{T_1}}
  \underbrace{\begin{bmatrix}
    \boldsymbol{Z}_{M_A/T_1} & \boldsymbol{0}_{M_A/T_1}     & \cdots & \boldsymbol{0}_{M_A/T_1} \\
    \boldsymbol{0}_{M_A/T_1} & \boldsymbol{Z}_{M_A/T_1} & \cdots &  \boldsymbol{0}_{M_A/T_1}       \\
    \vdots & \vdots & \ddots & \vdots\\
    \boldsymbol{0}_{M_A/T_1} & \boldsymbol{0}_{M_A/T_1} & \cdots & \boldsymbol{Z}_{M_A/T_1}
  \end{bmatrix}}_{M_A}
\end{equation}
where $\boldsymbol{Z}_{N}$ denotes an $N\times{N}$ matrix where only the entry on the upper-left corner is one and all the other entries are zero. We design $\boldsymbol{W}^{+}$ as
\begin{equation}
  \boldsymbol{W}^{+} = \gamma_N\sqrt{\frac{N_A}{T_3}}
  \underbrace{\begin{bmatrix}
    \boldsymbol{Z}_{N_A/T_3} & \boldsymbol{0}_{N_A/T_3}     & \cdots & \boldsymbol{0}_{N_A/T_3} \\
    \boldsymbol{0}_{N_A/T_3} & \boldsymbol{Z}_{N_A/T_3} & \cdots &  \boldsymbol{0}_{N_A/T_3}       \\
    \vdots & \vdots & \ddots & \vdots\\
    \boldsymbol{0}_{N_A/T_3} & \boldsymbol{0}_{N_A/T_3} & \cdots & \boldsymbol{Z}_{N_A/T_3}
  \end{bmatrix}}_{N_A}
\end{equation}

Then we analyze the largest entry of $\boldsymbol{r}_M$. The amplitude of $\boldsymbol{r}_M[k]$ can be derived by (\ref{rNk2}).
\begin{figure*}[ht]
\begin{align}\label{rNk2}
   \big|\boldsymbol{r}_M[k]\big| &= \big|\boldsymbol{\alpha}(M_A,-1+2(k-1)/K)^H \boldsymbol{F}_u \boldsymbol{F}_u^H \boldsymbol{\alpha} (M_A,\varphi_g)\big| \notag \\
     &= \gamma_M\sqrt{\frac{M_A}{T_1}} \Bigg|\boldsymbol{\alpha}(M_A,-1+2(k-1)/K)^H
  \begin{bmatrix}
    \boldsymbol{Z}_{M_A/T_1} & \boldsymbol{0}_{M_A/T_1}     & \cdots & \boldsymbol{0}_{M_A/T_1} \notag \\
    \boldsymbol{0}_{M_A/T_1} & \boldsymbol{Z}_{M_A/T_1} & \cdots &  \boldsymbol{0}_{M_A/T_1}      \notag \\
    \vdots & \vdots & \ddots & \vdots\\
    \boldsymbol{0}_{M_A/T_1} & \boldsymbol{0}_{M_A/T_1} & \cdots & \boldsymbol{Z}_{M_A/T_1}
  \end{bmatrix} \boldsymbol{\alpha} (M_A,\varphi_g)\Bigg| \\
     &= \frac{{\gamma_M }}{{M_A}}\sqrt{\frac{M_A}{T_1}} \Big| \sum_{i=1}^{T_1} e^{-j\pi(\varphi_g-(-1+2(k-1)/K))((i-1)M_A/T_1)} \Big| = \frac{{\gamma_M }}{{M_A}}\sqrt{\frac{M_A}{T_1}} \Big| \sum_{i=1}^{T_1} e^{-j\pi M_A/T_1(\varphi_g-(-1+2(k-1)/K))(i-1)} \Big| \notag \\
     &=\frac{{\gamma_M T_1}^2}{{M_A}} \sqrt{\frac{M_A}{T_1}}  \big|\boldsymbol{\alpha}(T_1,M_A(-1+2(k-1)/K)/T_1)^H \boldsymbol{\alpha} (T_1,\varphi_gM_A/T_1)\big|.
\end{align}
\hrulefill
\end{figure*}
\begin{figure*}[ht]
\begin{align}\label{rNk3}
   \big|\boldsymbol{r}_M[k]\big| &= \big|\boldsymbol{\alpha}(M_A,-1+2(k-1)/K)^H \boldsymbol{F}_u \boldsymbol{F}_u^H \boldsymbol{\alpha} (M_A,\varphi_g)\big| \notag \\
     &= \gamma_M\sqrt{\frac{M_A}{T_1}} \Big|\boldsymbol{\alpha}(M_A,-1+2(k-1)/K)^H \begin{bmatrix} \boldsymbol{I}_{T_1} & \boldsymbol{0}_{(T_1)\times{(M_A-T_1)}} \\ \boldsymbol{0}_{(M_A-T_1)\times{(T_1)}} & \boldsymbol{0}_{(M_A-T_1)\times{(M_A-T_1)}} \end{bmatrix} \boldsymbol{\alpha} (M_A,\varphi_g)\Big| \notag \\
     &= \frac{\gamma_M{T_1}^2}{{M_A}} \sqrt{\frac{M_A}{T_1}} \Big|\boldsymbol{\alpha}(T_1,-1+2(k-1)/K)^H \boldsymbol{\alpha} (T_1,\varphi_g)\Big|
\end{align}
\hrulefill
\end{figure*}
It is verified by (\ref{rNk2}) that the off-diagonal entries of $\boldsymbol{F}_u^{+}$ are all zero. We observe that
$\big|\boldsymbol{r}_M[k]\big|$ is essentially the inner product between a steering vector $\boldsymbol{\alpha}(T_1,M_A(-1+2(k-1)/K)/T_1)$ and a steering vector $\boldsymbol{\alpha} (T_1,\varphi_g M_A/T_1)$, indicating that $\big|\boldsymbol{r}_M[k]\big|$ is maximized when the angles of these two steering vectors are the closest, i.e.,
\begin{equation}\label{ClosestSteeringAngle}
  \widehat{k} = \arg\underset{1\leq{k}\leq{K}}{\min} \big| M_A \big(-1+2(k-1)/K-\varphi_g\big) / T_1 +2l \big|,~l \in\mathbb{Z}.
\end{equation}
Note that the angles of these two steering vectors may be out of $[-1,1]$ because $M_A/T_1>1$. Therefore we need to add the term of $2l$ in (\ref{ClosestSteeringAngle}) to guarantee that $(-1+2(k-1)/K-\varphi_g)M_A/T_1 + 2l$ is within $[-1,1]$. From (\ref{ClosestSteeringAngle}), we obtain
\begin{equation}\label{solutions_k_hat}
  \widehat{k}=\big\langle \big(-2 T_1 l / M_A+\varphi_g +1 \big) K/2    \big\rangle +1,~l \in\mathbb{Z}.
\end{equation}
It is seen that the number of solutions to (\ref{solutions_k_hat}) is $\lfloor M_A/T_1 \rfloor$.
Define $\widehat{\varphi}_g \triangleq -1+2(\widehat{k}-1)/K$ as an estimation of $\varphi_g$. We have
\begin{equation}\label{estiTheta}
\widehat{\varphi}_g= 2 \big\langle \big(-2 T_1 l / M_A+\varphi_g +1 \big) K / 2\big\rangle / K -1,~l \in\mathbb{Z}.
\end{equation}
Suppose $T_1 K/M_A$ to be an integer. Then (\ref{estiTheta}) can be rewritten as
\begin{equation}\label{estiTheta2}
\widehat{\varphi}_g= -2T_1 l / M_A + 2 \big \langle K(\varphi_g+1)/2 \big\rangle / K -1,~l \in\mathbb{Z}.
\end{equation}
where $2\langle(\varphi_g+1)K/2\rangle /K-1$ is essentially the quantization of $\varphi_g$ with resolution of $2/K$. It seen from (\ref{estiTheta2}) that $\widehat{\varphi}_g$ is periodic with $l$, indicating that the main lobe and the side lobes of $|\boldsymbol{r}_M|$ have the same envelope, which is illustrated in the third sub-figure of Fig.~\ref{FIG_1}.

Similarly, we define $\theta_g \triangleq \theta_{u,1}$ to ease the notation, and further define $\widehat{\theta}_g$ as an estimation of $\theta_g$, we have
\begin{equation}\label{estiVar2}
\widehat{\theta}_g= -2T_3 i / N_A + 2 \big \langle K(\theta_g+1) / 2 \big\rangle / K -1,~i \in\mathbb{Z}.
\end{equation}
It seen from (\ref{estiVar2}) that $\widehat{\theta}_g$ is periodic with $i$.

In order to eliminate the uncertainty of $l$ in (\ref{estiTheta2}) and $i$ in (\ref{estiVar2}), we resort to beam training based on codebook \cite{he2017codebook} to find the main lobes of $|\boldsymbol{r}_M|$ and $|\boldsymbol{r}_N|$, and then design $\boldsymbol{F}_u$ and $\boldsymbol{W}$ to estimate the AoA and AoD of the LOS path within the main lobe.

\begin{figure}[!t]
\centering
\includegraphics[width=60mm]{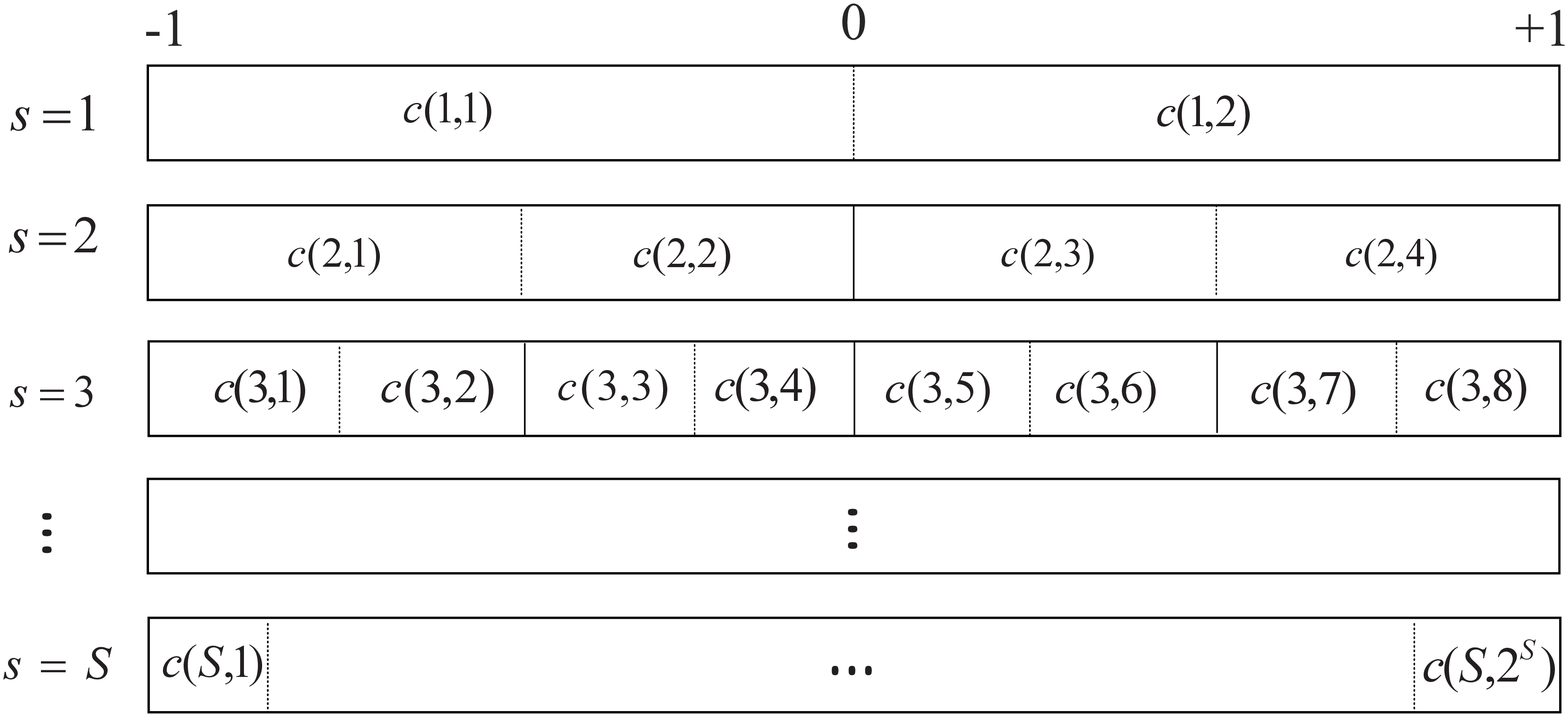}
\caption{The structure of hierarchical codebook.}
\label{FIG_3}
\end{figure}

\begin{figure}[!t]
\centering
\includegraphics[width=70mm]{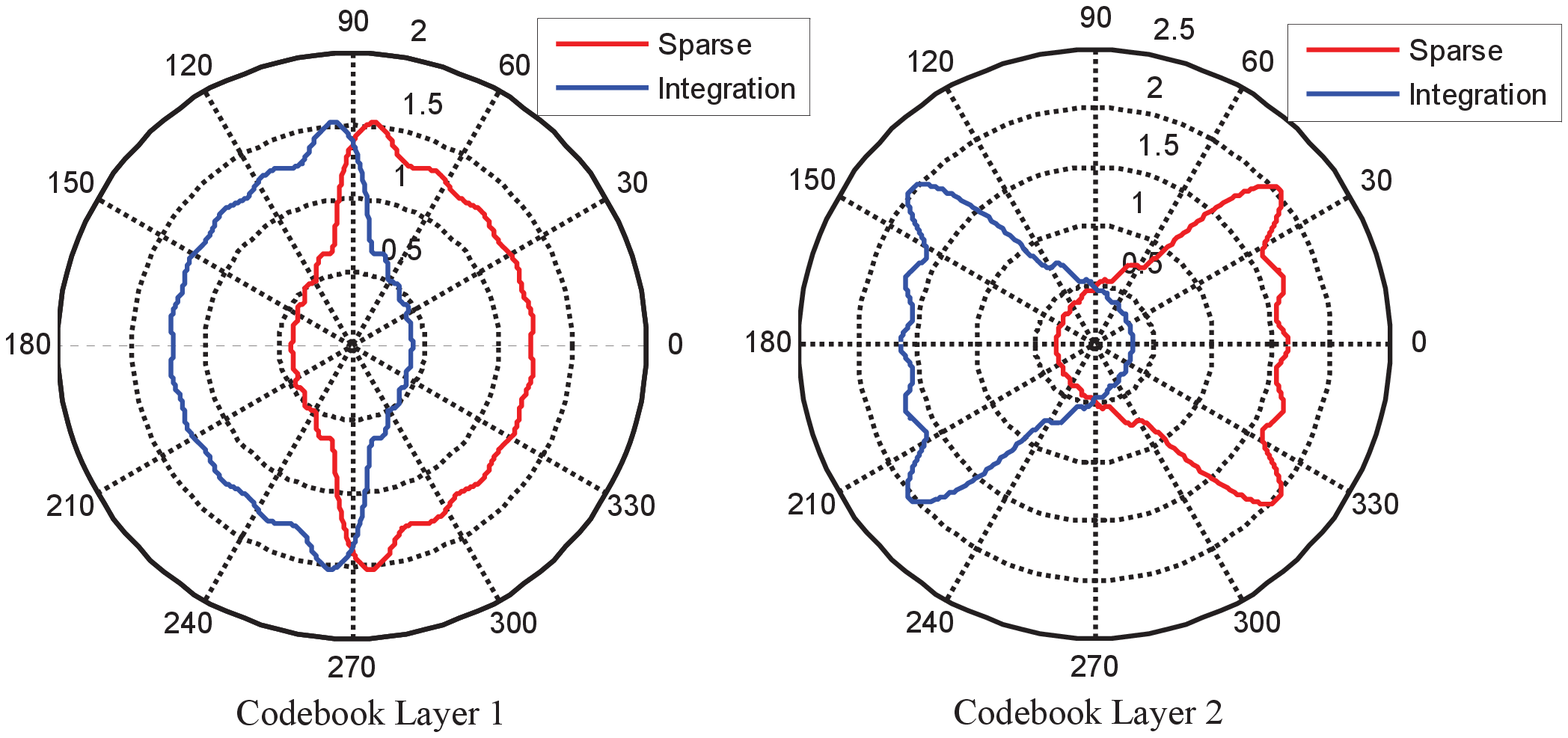}
\caption{Comparison of beam patterns designed by the integration-based method and the sparse-based algorithm.}
\label{FIG_4}
\end{figure}

\textbf{(i)} \textbf{Codebook Design and Beam Training}:
As shown in Fig.~\ref{FIG_3}, a typical  hierarchical codebook has $S$ layers, which satisfy $N=2^S$, where $N$ is the number of antennas. In the $s(s=1,2,\ldots,S)$th layer, there are $2^s$ codewords with the same beam width but different steering angles. The union of beam angle of all the codewords in each layer is $[-1, 1]$. Denote $\boldsymbol{c}(s,n)$ as the $n(n=1,2,\ldots,2^s)$th codeword in the $s$th layer, covering the beam angle
\begin{equation}
  \boldsymbol{\omega}(s,n) \triangleq [\omega_1(s,n),\omega_2(s,n)]
\end{equation}
where $\omega_1(s,n) \triangleq -1+(n-1)/2^{s-1}$ and $\omega_2(s,n) \triangleq -1+n/2^{s-1}$. Now we propose an integration-based codebook design method. The codeword $\boldsymbol{c}(s,n)$ is the integration of steering vectors $\boldsymbol{\alpha}(N,\theta)$ with $\theta$ from $\omega_1(s,n)$ to $\omega_2(s,n)$, which is expressed in (\ref{codebook}). In Fig.~\ref{FIG_4}, we compare the integration-based codebook design method with the existing sparse-based codebook design method \cite{alkhateeb2014channel}, with respect to the codewords in the first and second layer of the codebook. It shows the codewords designed by the integration-based method have the same beam pattern as those designed by the sparse-based method. However, our method has a closed-form expression, which is much easier for beam generation than the sparse-based method.

After transmitting the beams formed by the codewords of the $S$th layer, the AoA and AoD can be coarsely estimated within $[-1+2(n_W-3/2)/N_A,-1+2(n_W-1/2)/N_A$ and $[-1+2(n_F-3/2)/M_A,-1+2(n_F-1/2)/M_A]$, respectively, where $n_F \in \{1,2,\ldots,M_A\}$ and $n_W \in \{1,2,\ldots,N_A\}$ are the codeword indices of the $S$th layer. After finishing the search of the codebook, the resolution of the AoA and AoD is $2/N_A$ and $2/M_A$, respectively.

\begin{figure*}[hb]
\hrulefill
\begin{align}\label{codebook}
   \boldsymbol{c}(s,n) &= \int_{\omega_1(s,n)}^{\omega_2(s,n)} \boldsymbol{\alpha}(N,\theta) \textrm{d}\theta \notag \\
     &= \int_{\omega_1(s,n)}^{\omega_2(s,n)} \frac{1}{\sqrt{N}}\left[1,e^{-j\pi\theta},...,e^{-j\pi\theta(N-1)}\right]^{T} \textrm{d}\theta \notag \\
     &= \frac{1}{\sqrt{N}}\left[\frac{1}{2^{s-1}},\frac{j}{\pi}(e^{-j\pi\omega_2(s,n)}-e^{-j\pi\omega_1(s,n)}),...,\frac{j}{\pi}(e^{-j\pi(N-1)\omega_2(s,n)}-e^{-j\pi(N-1)\omega_1(s,n)})\right]^{T} \notag \\
     &= \frac{1}{\sqrt{N}}\left[\frac{1}{2^{s-1}},\frac{j}{\pi}(e^{-j\pi\omega_2(s,n)}(1-e^{j\pi/2^{s-1}}),...,\frac{j}{\pi(N-1)}(e^{-j\pi(N-1)\omega_2(s,n)}(1-e^{j\pi(N-1)/2^{s-1}}))\right]^{T}
\end{align}
\end{figure*}

\textbf{(ii)} \textbf{Complementary Channel Estimation}:
After the codebook training in \textbf{(i)}, the main lobes of $|\boldsymbol{r}_M|$ and $|\boldsymbol{r}_N|$ are identified. However, the estimation precision of the AoA and AoD is not satisfied. Therefore, we make complementary channel estimation after the codebook training. During the complementary channel estimation, limited pilot transmission with small pilot overhead is needed.

As illustrated in the third sub-figure of Fig.~\ref{FIG_1}, the width of the main lobe of $|\boldsymbol{r}_M|$ is $4/M_A$, which requires the period of $2T_1/M_A$ no smaller than $4/M_A$, resulting in $T_1\geq 2$. Similarly, we have $T_3=T_2N_R \geq 2$, resulting in $T_2\geq 1$. We set $T_1=2$ and $T_2=1$, meaning that we only need two different hybrid precoding matrices $\boldsymbol{F}_u$ and one hybrid combining matrix $\boldsymbol{W}$, where the overhead of pilot training is substantially reduced compared to the IA-based scheme.

Finally, we run the \textbf{second stage} of $\mathbf{Algorithm~\ref{alg2}}$ with $\Gamma_1 \triangleq -1+2(n_W-3/2)/N_A$, $\Gamma_2 \triangleq -1+2(n_W-1/2)/N_A$, $\Upsilon_1 \triangleq -1+2(n_F-3/2)/M_A$ and $\Upsilon_2 \triangleq -1+2(n_F-1/2)/M_A$. Note that we cannot run the \textbf{first stage} of $\mathbf{Algorithm~\ref{alg2}}$ to find the main lobe of $|\boldsymbol{r}_M|$ and $|\boldsymbol{r}_N|$, since the main lobe and the side lobes have the same envelope, as shown in the third sub-figure of Fig.~\ref{FIG_1}. We can only use the beam training based on codebook to find the main lobe.

\subsubsection{\bf CZO-based Beamspace Channel Estimation Scheme}
In this scheme, the nonzero diagonal entries of $\boldsymbol{F}_u^{+}$ and $\boldsymbol{W}^{+}$ are concentrated on the upper left corner of the matrix. We design $\boldsymbol{F}^{+}_u$ and $\boldsymbol{W}^{+}$ as
\begin{equation}
  \boldsymbol{F}^{+}_u = \gamma_M\sqrt{\frac{M_A}{T_1}} \underbrace{\begin{bmatrix} \boldsymbol{I}_{T_1} & \boldsymbol{0}_{T_1\times(M_A-T_1)} \\ \boldsymbol{0}_{(M_A-T_1)\times{T_1}} & \boldsymbol{0}_{(M_A-T_1)\times{(M_A-T_1)}} \end{bmatrix}}_{M_A},
\end{equation}
\begin{equation}
  \boldsymbol{W}^{+} = \gamma_N\sqrt{\frac{N_A}{T_3}} \underbrace{\begin{bmatrix} \boldsymbol{I}_{T_3} & \boldsymbol{0}_{T_3\times(N_A-T_3)} \\ \boldsymbol{0}_{(N_A-T_3)\times{T_3}} & \boldsymbol{0}_{(N_A-T_3)\times{(N_A-T_3)}} \end{bmatrix}}_{N_A}.
\end{equation}

Then we analyze the largest entry of $\boldsymbol{r}_M$. Similar to (\ref{rNk2}), the amplitude of $\boldsymbol{r}_M[k]$ can be derived by (\ref{rNk3}). We also observe that $\big|\boldsymbol{r}_M[k]\big|$ is the inner product between a steering vector $\boldsymbol{\alpha}(T_1,-1+2(k-1)/K))$ and a steering vector $\boldsymbol{\alpha} (T_1,\varphi_g)$, indicating that $\big|\boldsymbol{r}_M[k]\big|$ is maximized when the angles of these two steering vectors are the closest, i.e.,
\begin{equation}\label{ClosestSteeringAngle2}
\widehat{k} = \arg\underset{1\leq{k}\leq{K}}{\min} \big|(-1+2(k-1)/K) - \varphi_g\big|.
\end{equation}
Note that different with (\ref{ClosestSteeringAngle}), there is no term of $2l$ in (\ref{ClosestSteeringAngle2}), because both $-1+2(k-1)/K$ and $\varphi_g$ are within $[-1,1]$. From (\ref{ClosestSteeringAngle2}), we obtain
\begin{equation}
  \widehat{k}=\big\langle \big(\varphi_g +1 \big) K \big/ 2    \big\rangle +1.
\end{equation}
Define $\widehat{\varphi}_g \triangleq -1+2(\widehat{k}-1)/K$ as an estimation of $\varphi_g$. We have
\begin{equation}\label{Alg3estThete}
\widehat{\varphi}_g= 2\big\langle \big(\varphi_g +1 \big) K / 2 \big\rangle / K -1.
\end{equation}
where $2\langle(\varphi_g+1)K/2\rangle /K-1$ is essentially the quantization of $\varphi_g$ with resolution of $2/K$. Unlike (\ref{estiTheta}), $\widehat{\varphi}_g$ in (\ref{Alg3estThete}) is not periodic. Therefore, the envelope of the main lobe and the side lobe is different, as shown in the last sub-figure of Fig.~\ref{FIG_1}, where the beam training based on codebook to identify the main lobe is not necessary. We can directly employ $\mathbf{Algorithm~\ref{alg2}}$ to find the largest entry, which corresponds to the AoD and AoA of the LOS path.

\subsection{Comparisons}
Now we compare the proposed three schemes together with the HMC-based, JOINT-based, ECS-based, DCS-based and OCS-based channel estimation scheme in terms of computational complexity, estimation error and total time slots for channel training, which are summarized in Table~\ref{Talbe1}.

\begin{table}[!t]
\centering
\caption{Comparisons of different schemes.}
\label{Talbe1}
\begin{tabular}{p{0.8cm}p{1.5cm}p{1.6cm}p{1.3cm}p{1.3cm}}
\toprule
          &   Computational Complexity &   Estimation Error without Noise     &   Estimation Error with Noise      &   Total Time Slots\\
\midrule
IA  &  low   &   $>2/K$         &  small  &   flexible   \\
SZO &   low  &   $2/K$  &   small  &   fixed   \\
CZO  &  low  &   $2/K$  &   large  &   flexible  \\
DCS      &   high         &   $>2/K$           &   large  &   flexible  \\
OCS      &   high          &   $>2/K$          &   large  &  flexible \\
ECS      &   high          &   $>2/K$          &   large  &  flexible \\
HMC      &   low          &   depend on $T$          &   small  &  flexible \\
JOINT      &   low          &   depend on $T$          &   small  &  flexible \\
\bottomrule
\end{tabular}
\end{table}

\subsubsection{Computational Complexity}
As shown in (65), (67) and $\mathbf{Algorithm~\ref{alg1}}$, the proposed hybrid precoding and combining schemes are independent of the channel matrix, so we can design the hybrid precoding and combining matrix well before the transmission of pilot sequences. In this way, we do not need to consider the channel coherence time when designing the hybrid precoding and combing matrix. In other words, the computational complexity mainly comes from the search of the largest entry in $\mathbf{Algorithm~\ref{alg2}}$. For both the IA-based scheme and CZO-based scheme, in the first stage, we need to compute Frobenius norm of the $N_A \times 2$ matrix $\bar{\boldsymbol{R}}_{u,q}^v$ in (\ref{find_sq}) for $M_A-1$ times and Frobenius norm of the $2 \times M_A$ matrix $\bar{\boldsymbol{R}}_{u,p}^v$ in (\ref{find_sp}) for $N_A-1$ times, resulting in the complexity to be $\mathcal{O}( 4N_A(M_A-1) + 4M_A(N_A-1) )$. In the second stage, we use the trichotomy search to find the largest entry among $2K/N_A \times 2K/M_A$ entries. Since the length of current $\boldsymbol{\Gamma}$ and $\boldsymbol{\Upsilon}$ is $2/3$ of that of the previous $\boldsymbol{\Gamma}$ and $\boldsymbol{\Upsilon}$ in (\ref{updategam}), the number of iterations is $\log_{3/2}(2K/M_A)$. In the first $\log_{3/2}(2K/N_A)$ iterations, we compute four trichotomy points in each iteration. The entry on each point is the multiplication of three parts, including a $1\times N_A$ row vector of $\boldsymbol{D}(N_A,K)^H$, the $N_A\times M_A$ matrix $\boldsymbol{W}^H \boldsymbol{W} \boldsymbol{H}_u \boldsymbol{F}_u \boldsymbol{F}_u^H$, and an $M_A\times 1$ column vector of $\boldsymbol{D}(M_A,K)$, leading to the complexity to be $\mathcal{O}(8(\log_{3/2}(2K/N_A))(N_A+1)M_A)$. In the following $\log_{3/2}(2K/M_A)-\log_{3/2}(2K/N_A)$ iterations, since the AoA has been estimated, we only need to compute two trichotomy points in each iteration. The entry on each point is the multiplication of two parts, including a $1\times M_A$ row vector of $\boldsymbol{D}(N_A,K)^H \boldsymbol{W}^H \boldsymbol{W} \boldsymbol{H}_u \boldsymbol{F}_u \boldsymbol{F}_u^H$ and an $M_A\times 1$ column vector of $\boldsymbol{D}(M_A,K)$, leading to the complexity to be $\mathcal{O}(4(\log_{3/2}(2K/M_A)-\log_{3/2}(2K/N_A))M_A)$. Therefore the total computational complexity for both the IA-based scheme and CZO-based scheme is
\begin{align}
  & \mathcal{O}( 4N_A(M_A-1) + 4M_A(N_A-1) + 8(\log_{3/2}(2K/N_A))\notag \\ &(N_A +1)M_A  + 4(\log_{3/2}(2K/M_A)-\log_{3/2}(2K/N_A))M_A ).
\end{align}
For the SZO-based scheme, where only the second stage of $\mathbf{Algorithm~\ref{alg2}}$ is needed, we use the trichotomy search to find the largest entry among $K/N_A \times K/M_A$ entries, resulting in the total computational complexity to be
\begin{align}
  &\mathcal{O}( 8(\log_{3/2}(K/N_A))(N_A+1)M_A +\notag \\ & 4(\log_{3/2}(K/M_A)-\log_{3/2}(K/N_A))M_A ).
\end{align}
For the DCS-based channel estimation scheme~\cite{dai2016estimation}, where the main lobe is first searched and then the largest entry within the main lobe is further searched by the exhaustive search method, the computational complexity is
\begin{equation}
  \mathcal{O}( 2N_A(M_A-1) + 2M_A(N_A-1) + 2K^2(N_A+1)/N_A).
\end{equation}
For the ECS-based~\cite{li2015estimation} and OCS-based~\cite{alkhateeb2015compressed} channel estimation schemes, the largest entry is directly searched by the exhaustive search method, the computational complexity is
\begin{equation}
  \mathcal{O}( 2M_A(N_A+1)K^2).
\end{equation}
Since $K>N_A$ and $K>M_A$, the computational complexity of the IA-based scheme, SZO-based scheme and CZO-based scheme is much lower than that of the ECS-based, OCS-based and DCS-based channel estimation scheme. Moreover, for the HMC-based and JOINT-based schemes, since they use hierarchical codebook for beam training instead of computing multiplication in (\ref{Ruv}) to obtain $\boldsymbol{R}_u^v$, the computational complexity is almost zero.

\subsubsection{Estimation Error}
For four schemes including the IA-based scheme, ECS-based scheme, DCS-based scheme and OCS-based scheme, the estimation error of the AoA and AoD is larger than $2/K$ without noise. However, as shown in (\ref{estiTheta2}) and (\ref{Alg3estThete}), the estimation error of the AoA and AoD is $2/K$ without noise for both SZO-based and CZO-based scheme. For the HMC-based scheme and JOINT-based scheme, the precision of channel estimation relies on
the number of the layers in the hierarchical codebook. In fact, the number of time slots for training is proportional to the layers in the codebook. However, for the IA-based scheme, ECS-based scheme, DCS-based scheme, OCS-based scheme, SZO-based scheme and CZO-based scheme, $K$ can be set independent of the number of time slots for training.

As shown in the last sub-figure of Fig.~\ref{FIG_1} for the CZO-based scheme, the width of main lobe of $\boldsymbol{r}_M$ and $\boldsymbol{r}_N$ is $4/T_1$ and $4/T_3$, respectively, which are larger than that of the IA-based scheme and SZO-based scheme. Therefore, $\boldsymbol{r}_N$ and $\boldsymbol{r}_M$ in the CZO-based scheme are flatter than those of the IA-based scheme and SZO-based scheme, meaning that the CZO-based scheme has the largest estimation error among the three schemes with the same noise power.

\subsubsection{Total Time Slots for Channel Training}
For the SZO-based scheme where the beam training based on codebook is used, the number of time slots consumed by the beam training is $U(5\log_2M_A+2\log_2(N_A/M_A))$. Additionally, we need $2U$ time slots to transmit pilot sequences. Therefore, the number of total times slots for channel training is $U(5\log_2M_A+2\log_2(N_A/M_A)+2)$ for the SZO-based scheme. However, for the IA-based scheme, CZO-based scheme, ECS-based scheme, DCS-based scheme, OCS-based scheme, HMC-based scheme and JOINT-based scheme, the total time slots are flexible, depending on $T_1$ and $T_2$.

\section{Simulation Results}
Now we evaluate the performance of three proposed schemes. We consider uplink transmission of a multi-user mmWave massive MIMO system. The BS serving $U=4$ users has $N_A=64$ antennas and $N_R=4$ RF chains, while each user has $M_A=16$ antennas and $M_R=1$ RF chain. Suppose we use 6 bit and 4 bit digital phase shifters at the BS and users, respectively. The number of resolvable paths in mmWave channel is set to be $L_{u}=3$, while $g_{u,1}\thicksim\mathcal{CN}(0,1)$ and $g_{u,i}\thicksim\mathcal{CN}(0,0.01)$ for $i=2,3$. We use $K=1024$ over-sampling steering vectors. For SZO-based scheme, the number of total time slots for channel training is fixed to be $U(5\log_2M_A+2\log_2(N_A/M_A)+2)=144$. For the IA-based scheme, SZO-based scheme, ECS-based scheme, HMC-based scheme and JOINT-based scheme, the total time slots for channel training are flexible.

\begin{figure}[!t]
\centering
\includegraphics[width=95mm]{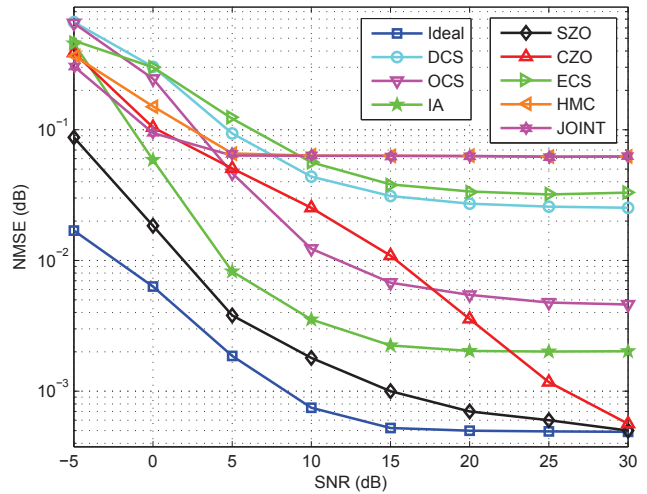}
\caption{Comparisons of NMSE for different SNR.}
\label{FIG_5}
\end{figure}

\begin{figure}[!t]
\centering
\includegraphics[width=95mm]{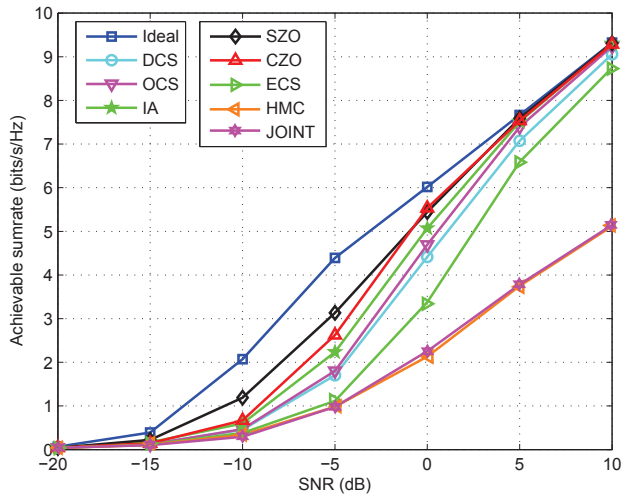}
\caption{Comparisons of sum-rate for different SNR.}
\label{FIG_6}
\end{figure}

\begin{figure}[!t]
\centering
\includegraphics[width=95mm]{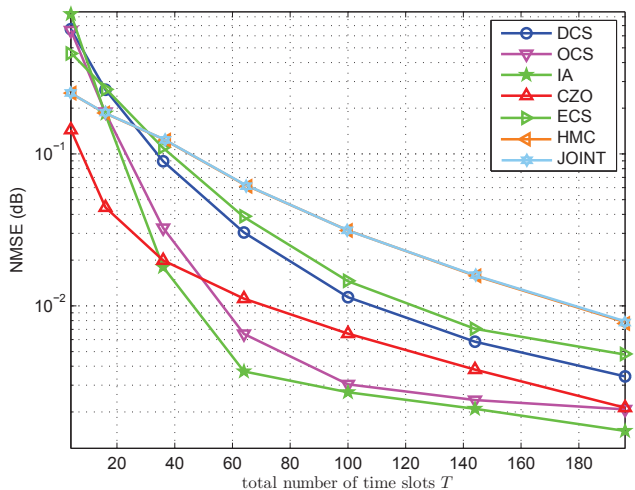}
\caption{Comparisons of NMSE for different number of total time slots.}
\label{FIG_7}
\end{figure}

\begin{figure}[!t]
\centering
\includegraphics[width=95mm]{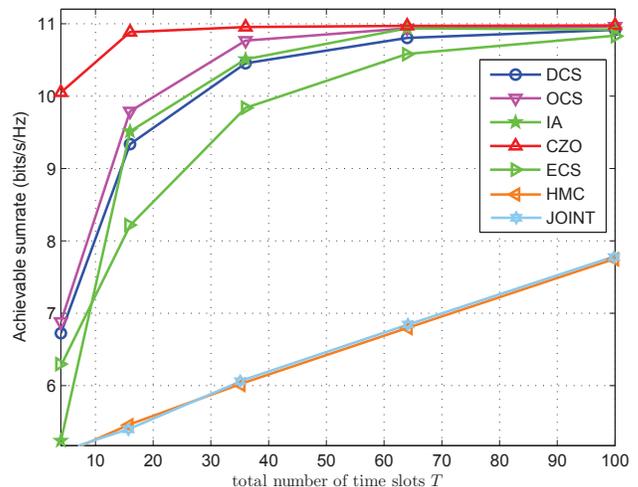}
\caption{Comparisons of sum-rate for different number of total time slots.}
\label{FIG_8}
\end{figure}

\begin{figure}[!t]
\centering
\includegraphics[width=95mm]{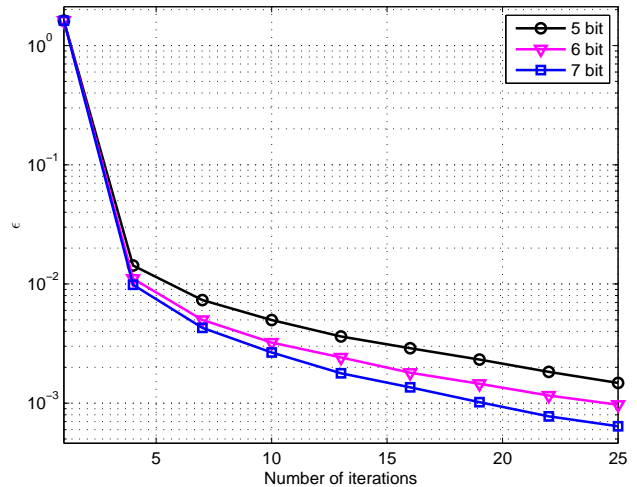}
\caption{Convergence of $\mathbf{Algorithm~\ref{alg1}}$.}
\label{FIG_9}
\end{figure}

As shown in Fig.~\ref{FIG_5}, we compare the channel estimation performance in terms of normalized mean square error (NMSE) for the proposed three schemes, the DCS-based scheme~\cite{dai2016estimation} and the OCS-based scheme~\cite{alkhateeb2015compressed} with different SNR. The NMSE is defined as $\frac{1}{U} \operatorname{E}\left\{ \sum_{u=1}^{U}\sqrt{(\hat{\theta}_{u,1}-\theta_{u,1})^2+(\hat{\varphi}_{u,1}-\varphi_{u,1})^2} \right\}$, which reflects the estimation accuracy of AoA and AoD. We set $T_1=4$ and $T_2=4$. Then $T_3=T_2N_R=16$. In order to make fair comparisons, we set the total time slots of the DCS-based scheme, OCS-based scheme, ECS-based scheme, HMC-based scheme and JOINT-based scheme the same as the proposed schemes. The number of total time slots for pilot training is fairly set to be $UT_1T_2=64$. Given $T_1T_2$, the number of codewords in the last layer of  hierarchical codebook is $2^{\langle T_1T_2/5\rangle}=8$. We also include the ideal case where $\boldsymbol{W}^H \boldsymbol{W}=\gamma_N\boldsymbol{I}_{N_A}$ and $\boldsymbol{F}_u \boldsymbol{F}_u^H=\gamma_M \boldsymbol{I}_{M_A}$ for comparisons.

It is observed from Fig.~\ref{FIG_5} that both the IA-based scheme and SZO-based scheme outperform the DCS-based,  OCS-based, ECS-based, HMC-based and JOINT-based schemes. At SNR of 15dB, the IA-based scheme has 92.8\%, 66.9\%, 94.2\%, 96.5\% and 96.5\% performance improvement compared with the DCS-based, OCS-based, ECS-based, HMC-based and JOINT-based schemes, respectively, while the SZO-based scheme has 96.8\%, 85.2\%, 97.4\%, 98.4\% and 98.4\%  improvement compared with the DCS-based, OCS-based, ECS-based, HMC-based and JOINT-based schemes, respectively. The reason for the performance of the DCS-based and OCS-based schemes is that both the DCS-based and OCS-based schemes employ random precoding and random combining matrix, which are not optimal. Although both the DCS-based and OCS-based schemes outperform the CZO-based scheme in low SNR region, at high SNR region such as SNR of 20dB, the CZO-based scheme has 86.8\% and 34.5\% improvement compared with the DCS-based and OCS-based schemes, respectively. The reason is that the width of the main lobe in CZO-based scheme is larger than DCS-based and OCS-based scheme. Besides, both the SZO-based scheme and CZO-based scheme can approach the ideal performance as SNR increases, which shows the superiority of SZO-based scheme and CZO-based scheme in their small estimation error. The reason for the unsatisfactory performance of the ECS-based scheme is that the ECS-based scheme does not consider the power leakage due to the limited beamspace resolution. Therefore the ideal sparse property of beamspace channel is impaired. The reason for the unsatisfactory performance of the HMC-based and JOINT-based schemes is the number of codewords in the last layer of hierarchical codebook is only 8 and  much smaller than $K=1024$, where the total time slots for training are set to be the same with other schemes for fair comparisons. In addition, compared with the CZO-based scheme, the IA-based scheme has smaller estimation error in the low SNR region. The IA-based scheme can achieve better NMSE performance than the CZO-based scheme when SNR $<$ 20 dB, which means we can use the IA-based scheme for better NMSE performance when SNR $<$ 20dB.

As shown in Fig.~\ref{FIG_6}, we compare the sum-rate for the proposed three schemes, the DCS-based, OCS-based, ECS-based, HMC-based and JOINT-based schemes. It is seen that the proposed three schemes achieve better performance than the other schemes, especially in low SNR region. At SNR of 10dB, the performance gap between the proposed three schemes and the ideal case is around 1dB.

As shown in Fig.~\ref{FIG_7}, we compare the channel estimation performance in terms of NMSE for different schemes with different number of total time slots for channel training, which is $T=UT_1 T_2$. Since the number of total time slots for channel training is fixed to be 144 for the SZO-based scheme, SZO-based scheme is not included for comparison. We fix SNR to be 15dB. Given $T$, we set $T_1=T_2=\sqrt{T/U}$. It is seen that when $T$ is small, the CZO-based scheme performs the best; when $T$ is large, the IA-based scheme performs the best. When $T=16$, the CZO-based scheme has 78.2\%, 77.8\%, 83.5\%, 76.0\% and 76.1\% performance improvement compared with the DCS-based, OCS-based, ECS-based, HMC-based and JOINT-based schemes, respectively. When $T=100$, the IA-based scheme has 76.3\%, 11.2\%, 81.4\%, 91.4\% and 91.5\% improvement compared with the DCS-based, OCS-based, ECS-based, HMC-based and JOINT-based schemes, respectively. In the ideal case that both $\boldsymbol{W}^H \boldsymbol{W}=\gamma_N\boldsymbol{I}_{N_A}$ and $\boldsymbol{F}_u \boldsymbol{F}_u^H=\gamma_M\boldsymbol{I}_{M_A}$ can be achieved, the number of total time slots for channel training is $UM_AN_A=4096$. It is seen from Fig.~\ref{FIG_7} that with substantially reduced training overhead, i.e., $T\ll 4096$, satisfactory performance can almost be achieved, e.g., 0.0027 of NMSE with $T=100$ for the IA-based scheme. The number of total time slots for channel training is flexible and fixed for the IA-based scheme and the SZO-based scheme, respectively, which indicates the flexibility of the IA-based scheme. To achieve the NMSE of 0.0027, $T=100$ is enough for the IA-based scheme, while $T=144$ is fixed for the SZO-based scheme.

As shown in Fig.~\ref{FIG_8}, we compare the sum-rate for different number of total time slots for channel training. We fix SNR to be 15dB. We observe that the CZO-based scheme can achieve better performance than the DCS-based, OCS-based, ECS-based, HMC-based and JOINT-based schemes when $T$ is small. When $T$ is close to 100, the sum-rate of the IA-based scheme and CZO-based scheme is almost invariant, indicating that $T=100$ is enough to achieve the maximal sum-rate.

As shown in Fig.~\ref{FIG_9}, we verify the convergence of $\mathbf{Algorithm~\ref{alg1}}$. Suppose we use 5, 6 and 7 bit digital phase shifters at the BS, respectively. It is seen that $\epsilon$ in (\ref{epsilon}) decreases rapidly as the number of iterations grows, and $\epsilon$ decreases faster with higher resolution of digital phase shifters. Using 6 bit digital phase shifters at the BS, $\epsilon$ is smaller than $10^{-3}$ when the number of iterations is larger than 25, which means 25 iterations is enough for $\delta=10^{-3}$.

\section{Conclusions}
This paper has investigated beamspace channel estimation for multi-user mmWave massive MIMO system. A framework of beamspace channel estimation has been proposed. Then based on the this framework, three channel estimation schemes have been proposed. These schemes together with the existing channel estimation schemes have been compared in terms of computational complexity, estimation error and total time slots for channel training. Simulation results have shown that the proposed schemes outperform the existing schemes and can approach the performance of the ideal case with substantially reduced training overhead. Future work will focus on the design of hybrid precoding and hybrid combining for multi-user data transmission regarding the energy efficiency and the theoretical proof of the convergence of iterative hybrid precoding and combining design.

\bibliographystyle{IEEEtran}
\bibliography{IEEEabrv,IEEEexample}

\end{document}